\documentclass[9pt,journal]{IEEEtran}

\usepackage{amsmath,epsfig}
\usepackage{color}

\usepackage{algorithm}
\usepackage{caption}
\usepackage{lipsum}
\usepackage{algorithmic}

\usepackage{setspace}

\usepackage{mathtools}
\DeclarePairedDelimiter{\ceil}{\lceil}{\rceil}

\ifCLASSINFOpdf

\else

\fi

\begin{document}
\onecolumn

\title{Graph-based representation for multiview image coding}

\author{Thomas~Maugey,~\IEEEmembership{Member,~IEEE,} Antonio Ortega,~\IEEEmembership{Fellow Member,~IEEE,}
        and~Pascal~Frossard,~\IEEEmembership{Senior Member~IEEE}
\thanks{Thomas Maugey and Pascal Frossard are with the Signal Processing Laboratory (LTS4), Institute of Electrical Engineering, \'Ecole Polytechnique F\'ed\'erale de Lausanne (EPFL), 1015 Lausanne, Switzerland (e-mail: thomas.maugey@epfl.ch;
pascal.frossard@epfl.ch).}
\thanks{Antonio Ortega is with the Department of Electrical Engineering, University
of Southern California, Los Angeles, CA 90089 USA (e-mail: antonio.ortega@sipi.usc.edu)}
}

\maketitle

\begin{abstract}
In this paper, we propose a new representation for multiview image sets. Our approach relies on graphs to describe geometry information in a compact and controllable way. The links of the graph connect pixels in different images and describe the proximity between pixels in the 3D space. These connections are dependent on the geometry of the scene and provide the right amount of information that is necessary for coding and reconstructing multiple views. This multiview image representation is very compact and adapts the transmitted geometry information as a function of the complexity of the prediction performed at the decoder side. To achieve this, our GBR adapts the accuracy of the geometry representation, in contrast with depth coding, which directly compresses with losses the original geometry signal. We present the principles of this graph-based representation (GBR) and we build a complete prototype coding scheme for multiview images. Experimental results demonstrate the potential of this new representation as compared to a depth-based approach. GBR can achieve a gain of $2$ dB in reconstructed quality over depth-based schemes operating at similar rates.
\end{abstract}


\IEEEpeerreviewmaketitle

\section{Introduction}

Multiview image coding has received considerable attention in recent years. In particular, hardware technologies for the capture and the rendering of multiview content have improved significantly. For example, depth sensors and autostereoscopic displays have become popular in the past years \cite{Alenya_G_2011_ieeesj}. This has led to novel immersive applications and thus to more challenges for the research community. One of the main open questions in multiview data processing resides in the design of representation methods for multiview data \cite{Shum_HY_2003_tcsvt_sur_ibrct, Muller_K_2011_pieee_tdv_rudm, Salvador_J_2013_mul_vrbfmcsr}, where the challenge is to describe the scene content in a compact form that is robust to lossy compression. Many  approaches have been studied in the literature such as the multiview format \cite{jmvm}, light fields \cite{Chai_JX_20000_psiggraph_ple_s} or even mesh-based techniques \cite{ Kim_SY_2007_picip_mes_bdctdvuhdm}. All these representations contain two types of data. On the one hand, the color or luminance information, which is classically described by 2D images. On the other hand, the geometry information describes the scene's 3D characteristics, represented by 3D coordinates, depth maps or disparity vectors\footnote{Note that no explicit scene geometry information is transmitted in the multiview case.}. 
Effective representation, coding and processing of multiview data rely on the proper manipulation of these two types of information, \emph{i.e.}, luminance and geometry.

\begin{figure}[t]
\centering
		\includegraphics[width=0.5\linewidth]{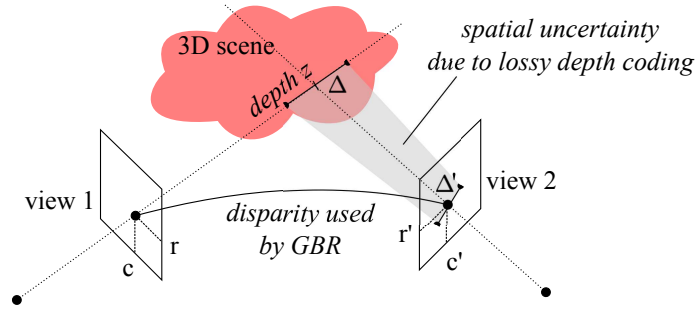}
	\caption{Pixel $(r,c)$ in view $1$ is associated to pixel $(r',c')$ in view $2$, given its geometry (depth value $z$). An uncertainty $\Delta$ about this pixel's depth leads to a spatial inaccuracy $\Delta'$ in view $2$. This basic observation is the origin of the main drawbacks of depth-based representations. In contrast, our GBR uses disparity values which are controlled lossy versions of depth values.}
	\label{fig:DQ}
\end{figure}

Since depth signals can  be efficiently captured due to the advent of new sensor devices, the multiview plus depth (MVD) \cite{Merkle_P_2007_picip_mvvpdrc} format has become very popular in recent years. Depth information allows us to build a reliable  estimation of scene geometry. With this information, encoders are able to extract the correlations between views \cite{Yea_S_2009_jspic_vie_spmvc}, and decoders can synthesize virtual views \cite{Tian_D_2009_pspie_vie_sttdv}. Many recent multiview video coders rely on depth signals to enhance their coding performance \cite{Muller_K_2013_tip_tdh_evcmvvdd}. However, the representation of geometry with depth data has one main drawback: if lossy compression is applied to depth, as done in classical coders, the induced error  makes it difficult to control the quality of synthesized viewpoint.  This is the case even if depth gives a good estimation of 3D scene geometry.  More specifically, uncertainty $\Delta$ in the depth value  (due to quantization for example) leads to a spatial uncertainty $\Delta'$ when determining the correspondence between pixels in neighboring views. This is illustrated in Fig~\ref{fig:DQ}.  Proper modeling of the impact of quantization on rendered view quality is in general difficult, even though it is crucial for solving classical problems such as rate allocation between depth and color signals in order to maximize the quality of the reconstructed views.

Note that, by nature, depth information represents the geometry of one view without considering any information about the predicted viewpoints. For example, the raw depth maps generally have too much precision given the view predictions they are supposed to perform. Instead of directly coding the raw depth maps with hard to control losses, a more efficient approach may consist of building a representation that captures only the information needed for the required view predictions, and then to perform a lossless coding of this new geometry signal\footnote{Note that this would be a lossy representation since only the information needed for rendering will be transmitted.}. This approach is similar to one based on the disparity vectors, even though these have a block precision in the current standards where they are employed. Hence, we investigate in this paper a solution for building ``just enough" geometry information for coding a given set of views. The proposed approach considers only integer disparities that are obtained after a rounding operation on the float disparity values derived from the depth maps. This geometry information can be viewed as a dense disparity map, which explicitly contains the information to link pixels in different views.

\begin{figure*}[t]
\centering
		\includegraphics[width=1\linewidth]{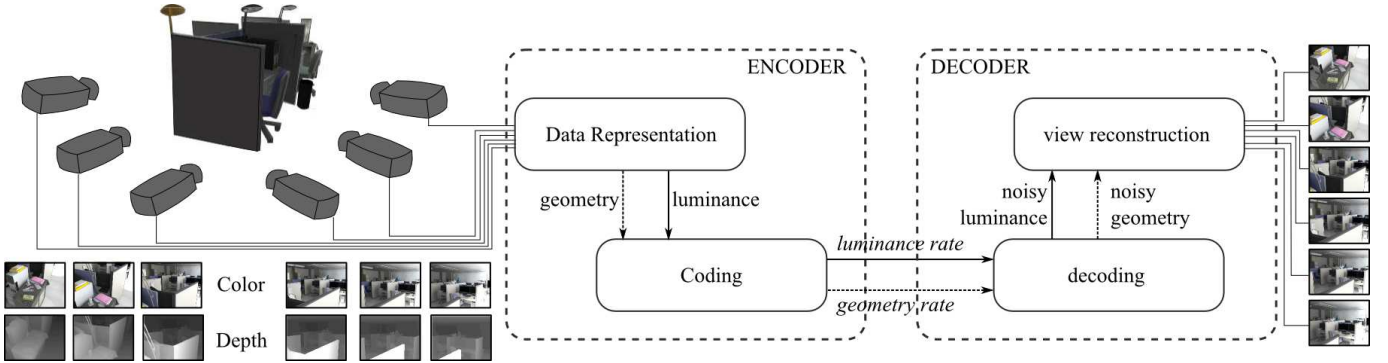}
	\caption{Our goal is to represent and code $N$ viewpoints with ground-truth color and depth informations, with a low rate and high decoding quality.}
	\label{fig:framework}
\end{figure*}

We propose a new geometry representation format based on graph structures, called Graph-Based Representation (GBR), where the geometry of the scene is represented as connections between corresponding pixels in different images. In this notation, two connected pixels are neighbors in the 3D scene. As shown in Fig.~\ref{fig:DQ}, they are derived from dense disparity maps and provide just enough geometry information to predict another viewpoint. 
In other words, before losslessly coding the geometry signal, GBR drastically simplifies it as a  function of how it will be used at the decoder. This ``use-aware" geometry compression allows us to control the error due to coding.
While GBR representations offer a very generic format, we focus our study on the scenario where $N$ views (color and depth), acquired at the encoder, are transmitted to a decoder that reconstructs the luminance images for all the $N$ views. This scheme is illustrated in Fig~\ref{fig:framework}. Throughout this paper, we compare our GBR solution, where geometry is represented by connections between pixels, to approaches where geometry is described by depth. We outline the importance of proper control of coding errors and show that it leads to a better view reconstruction quality.

In more detail, the GBR is constructed as follows. The first image in the set is represented by its color information. Then the GBR represents the \emph{new} pixels of image $2$ (\emph{i.e.}, pixels that are not present in image $1$, such as disoccluded pixels) and link them to their neighbors in image $1$ which correspond to the same 3D points in the scene. The same approach is repeated for all views (or a subset of them as explained later) until the $N^{th}$ view is reached. Hence, the resulting representation describes 3D points of the scene \emph{once and only once}, \emph{i.e.,} the first time they are captured by one of the cameras, and links them through the different views in the graph. We build on \cite{Maugey_T_2013_picassp_gra_brcmg, Maugey_T_2013_ivmsp_mul_icugba} where the basic concept of GBR has been introduced and here we extend that work by designing a complete scheme, where luminance information is coded along with the graph information. Moreover, we take into account the errors in the connections and introduce residual images that allows us to correct minor geometrical distortions. Our GBR-based multiview coding scheme thus has to transmit one reference image, the graph connections providing the geometrical information, the luminance signal of new pixels of every viewpoint and finally some residual images. As a first prototype implementation, we make use of off the shelf tools: JPEG2000 for image and residual, and arithmetic coding for the graph.

Throughout this paper, we compare our approach to a simplified depth-based scheme. Rather than using the most recent standards, which apply depth-based intra prediction to each block, we build an hybrid coding scheme where depth-based prediction is used for the whole image, and prediction residuals are transmitted. Image by image, the current view (color and depth images) is predicted using the previous view and the corresponding depth, and  then residuals for luminance and depth signals are sent. The residuals for luminance and depth correct the prediction errors and complete the information in the disoccluded regions. The reconstructed images eventually serves to build an estimate of the images from the following viewpoints.  This simplified approach provides a more direct way of comparing depth-based techniques to our GBR approach, since in both cases, the encoder is required to use geometry-based prediction for all images, except for the first one.
For depth image compression, we also use JPEG2000, which helps us to highlight the difficulties due to ``blind" geometry compression.
 In this paper we provide a proof of concept implementation of our GBR, rather than optimized RD results.  Our experiments nevertheless demonstrate that our GBR representation leads to an easier control of geometry compression artifacts than depth-based representation signals, leading to a better reconstruction quality.

Our GBR thus constitutes a promising alternative to depth-based representations that face the problem of geometry inaccuracies due to lossy compression of depth information. A number of approaches have been proposed recently to address this problem. In particular, some recent methods aim at improving rate-distortion performance of standard compression tools when applied to depth information.  For that purpose, models of the error in geometry estimation due to traditional lossy compression of depth have been studied. In \cite{Morvan_Y_2007_ppcs_joi_dtbamvvc}, the optimization is done by experimentally simulating some practical RD points and choosing the best one. The minimization is done with a multi-resolution full search. In some other works, a rate-distortion (RD) model is developed \cite{Kim_WS_2009_picip_dep_mdavrdc}. For example, in \cite{Wang_Q_2012_tcsvt_fre_vvcrda}, the RD model is estimated region-by-region, corresponding to the different objects of the scene. In \cite{Cheung_G_2011_tip_dep_bamicdibr}, the RD analysis relies on some complex models for the image textures. In \cite{Rajei_B_2012_at_rat_damcdibrf},  wavelet properties are used to separate the different components of the scene and to analyze object by object the consequence of inaccuracies in their depth values. Note that, regardless of the chosen RD model,  optimization remains complex and strongly dependent on scene content and camera settings (baseline, geometry complexity, etc.).

Instead of optimizing standard codecs as described above, another solution is to develop alternative coding tools for depth maps in order to enhance the control of depth compression. The main observation in these approaches is that depth maps have sharp edges but very smooth textures. The goal of these coding tools is to preserve the sharpness of the edges, while spending few bits on the flat or smooth parts. Examples of tools that have been proposed, include meshes \cite{Kim_SY_2007_picip_mes_bdctdvuhdm}, new block formats \cite{Liu_S_2011_tb_new_dctucv}, graph-based transforms \cite{Cheung_G_2011_mmsp_dep_mcugbttds} and coding of depth edges \cite{Daribo_I_2012_picip_ari_ecassmpdvc}. These tools indeed permit to increase the performance of the depth-based coding schemes. However, they do not provide a better understanding of the effect of depth compression on rendering. 

The same objective of reducing geometry inaccuracies has also be targeted by works that investigate alternative representations for multiview data.
Similarly to GBR, the layered depth image (LDI) representation \cite{Gelman_A_2012_tip_mul_icudloba,Takyar_U_2013_spl_ext_ldirmn}  avoids the inter-view redundancies in the signal description. More precisely, in both GBR and LDI, the 3D points of the scene are represented once and only once, which is not the case for light field, multiview or depth-based representations. In LDI \cite{Gelman_A_2012_tip_mul_icudloba, Takyar_U_2013_spl_ext_ldirmn}, the pixels of multiple viewpoints are projected onto a single view. The redundant pixels are discarded and the new ones (\emph{i.e.}, the ones occluded on this reference view) are added in an additional layer. This very promising representation has however the drawback of being dependent on the depth signal. Indeed, the LDI describes also the depth values in multiple layers. They are necessary at the decoder side for retrieving the viewpoints. Thus, the problem of controlling the error due to depth compression, mentioned for multiview-plus-depth format, still arises in LDI. A better control of these inaccuracies is achieved with GBR.

The rest of this paper is organized as follows. In Section~\ref{sec:GBR}, we present our GBR solution by introducing in detail the graph construction process and the view reconstruction technique. We then present the complete coding scheme for the transmission of multiview data with our GBR representation (Section~\ref{sec:coding}). Finally, in Section~\ref{sec:exp}, we present various experiments to compare the depth-based scheme and the GBR approach and we show the benefit of representing geometry with graphs. 

\section{Graph-based geometry representation}\label{sec:GBR}

\subsection{Multiview image data}

Let us consider a scene captured by $N$ cameras with the same resolution and focal length $f$. The $n$-th image is denoted by $I_n$, with $1\leq n \leq N$, where $I_n(r,c)$ is the pixel at row $r$ and column $c$. We consider translation between cameras, and we assume that the views are rectified. In other words, the geometrical correlation between the views $I_n$ depends on horizontal components. We also work under the Lambertian assumption, which states that each 3D point of the scene has the same luminosity when viewed from every possible viewpoint. We assume a depth image, $Z_n$, is available at the encoder for every viewpoints, $I_n$, as illustrated in Fig.~\ref{fig:framework}. Since the images are rectified, the relation between the depth $z$ and the disparity $d$ for two camera images is given by $d = \frac{f\delta}{z}$, where $\delta$ is the distance between the two cameras. In what follows, the geometry information is given by disparity values that are computed from the depth maps $Z_n$ and the camera parameters. Our goal is to design a compact multiview representation of these $N$ camera images that offers control of the geometry information accuracy.

\subsection{Geometrical structure representation}

\begin{figure}[t]
	\centering	\includegraphics[width=0.5\linewidth]{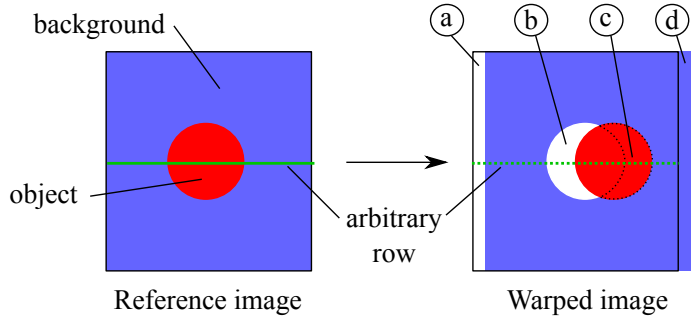}
	\caption{Illustration of camera translation for a simple scene with a uniform background, and one foreground object. Types of pixels in depth-based inter-view image warping: pixels can be a) appearing, b) disoccluded, c) occluded  and d) disappearing. The green plain line is an arbitrary row in the reference image and the dashed line is the corresponding row in the target image.}
	\label{fig:typesOfconnextions}
\end{figure}

Before introducing our new data representation in detail, we analyze  the  effect of camera translation on the image content. Let us consider two images $I_n$ and $I_{n+1}$ captured by cameras that are separated by a distance  $\delta$. Since we  consider only full pixel displacements, the geometrical correlation between  pixels in these two images takes the form of $I_{n+1}(r,c) = I_{n}(r,c+d)$, where $d$ is a disparity value. When  this relation holds, pixels in certain regions in image $I_{n+1}$ can be directly associated to pixels in corresponding regions in $I_{n}$. These correspond to the elements of the scene that are visible in both images. Alternatively, the elements that are visible only from one viewpoint are often designed under the general name of occlusions, even if their appearance is not only due to object occlusions.  More precisely, we can categorize these pixels that are present or absent only in one image, into four different types as  illustrated in Fig.~\ref{fig:typesOfconnextions}.  First, a new part of the scene appears in the camera because of camera translation. It usually comes from the right or left (depending on translation direction) and the new pixels are not related to object occlusions. They are called \emph{appearing} pixels. During camera translation, foreground objects move faster than the background. As a result, some background pixels may appear behind objects and are thus called \emph{disoccluded} pixels. Conversely, some background pixels may become hidden by a foreground object. These are called the \emph{occluded} pixels. Finally, some pixels disappear in the viewpoint change, and they are called \emph{disappearing} pixels. 

We illustrate these different types of pixels and consider a row of the target image in Fig.~\ref{fig:typesOfconnextions}. Starting from the left border, we notice that the row first contains several appearing pixels, and then some pixels of the reference image. Then, the row presents some disoccluded pixels before coming back to pixels of the reference image. After that, the row contains occluded pixels that correspond to a jump between pixels in the reference image. The rest of the row refers to the reference image until a series of disappearing pixels are depicted at the end of the row. We want now to describe the pixels in this target row in the second view by maximizing references to elements from the corresponding row in the first view. This can be achieved by navigating between the reference image and the ``new" pixels of the target image.  This navigation can be guided by connections between corresponding pixels in both views. We thus propose to construct a graph that is exactly made of these connections.  This graph is derived from the depth information and the number of connections  varies linearly with the number of foreground objects in the image. Similarly, the size of these connections  evolves linearly with the distance between cameras and object disparities. A more formal description of the graph construction method is given next.

\subsection{Graph construction}
\begin{figure*}[t]
	\centering	\includegraphics[width=1\linewidth]{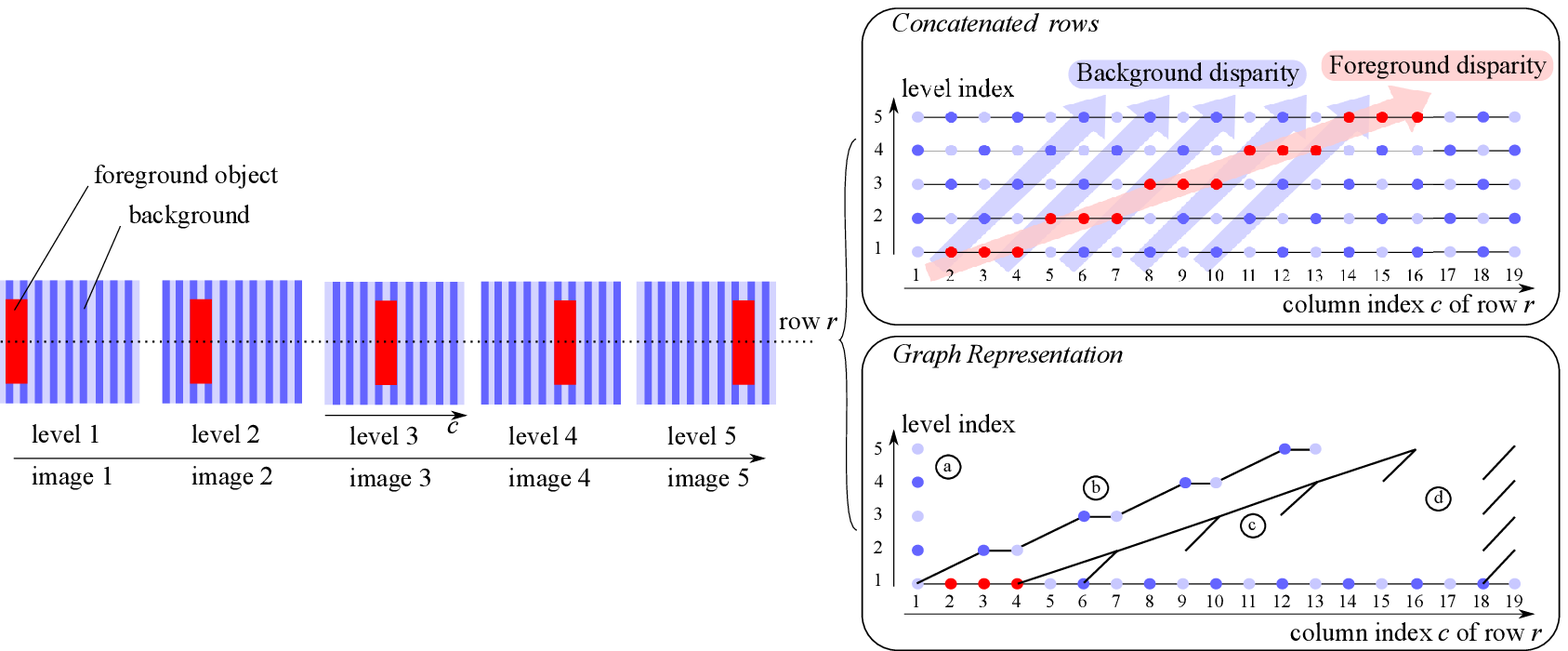}
	\caption{Graph construction example: the blue texture background has a disparity of $1$ at each view and the red rectangle foreground has a disparity of $3$ for each view. This example graph contains all different types of pixels: a) appearing, b) disoccluded, c) occluded  and d) disappearing.}
	\label{fig:intuitiveexample}
	\vspace{0.5cm}
		\centering	\includegraphics[width=1\linewidth]{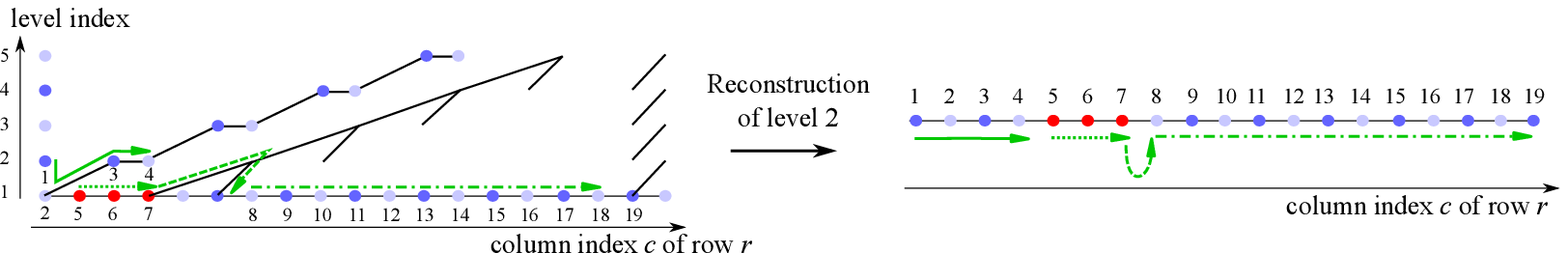}
	\caption{Reconstruction of the view $2$ with the toy example of Fig.~\ref{fig:intuitiveexample}. The green arrows indicates the graph exploration order for view reconstruction.}
	\label{fig:reconstructionexample}
\end{figure*}

The proposed graph representation intends to avoid redundancies in the color information (\emph{i.e.}, only ``new" pixels are described) and additionally to offer an intuitive description of the geometry information with links between corresponding pixels in different views. Generally, a graph with $N$ levels describes $1$ reference image and $N-1$ predicted ones and is constructed based on the depth maps $Z_n$, $1 \leq n \leq N-1$.
Since the object displacement is only horizontal, we consider that the graph construction is independent for each image row. 
For each row, the graph is made of two components, which are described by two matrices $\Gamma$ and $\Lambda$ of  size $L\times W$, where $L$ is the number of levels (\emph{i.e.,} the number of images encoded by the graph) and $W$ is the image width in pixels. These two matrices respectively gather color and geometry information for all pixels in that row across all images. The color values in row $r$ are given by $\Gamma_r = [ \gamma^r_{i,j}]_{i \leq L, j \leq W}$ and the connections in the same row are given by $\Lambda_r = [ \lambda^r_{i,j}]_{i \leq L, j \leq W}$. In the graph construction, both matrices are initialized to $0$, which means ``no connection" and ``no color value" respectively.

We now describe in details the construction of the graph. We show in Fig.~\ref{fig:intuitiveexample} a graph construction example, with $5$ levels that correspond to $1$ reference view and $4$ synthesized views. For the sake of clarity, we first describe in detail the graph construction of an arbitrary row $r$ by considering only one predicted view $I_{2}$, one reference view $I_1$ and its associated depth map $Z_1$. The first level corresponds to the reference view, and thus $\gamma^r_{1,j} =  I_1(r,j)$ for all $j\leq W$. The connections then indicate the relation between the pixels in the current level and those in the next one. 

Then, the connection values $\lambda^r_{1,j}$ and the color values $\gamma^r_{2,j}$ are assigned based on the following principles:
\begin{itemize}
\item The pixels intensities are represented in the level where they appear first, which means that the second level only contains pixels that are not present in the reference image. 
\item The connexions $\lambda^r_{1,j}$ simply consists in linking these ``new" pixels to the position of their neighbor in the previous level. More precisely, a new pixel represented at a level larger than $1$ is hidden by a foreground object in the previous views. If this foreground object was not in the scene, the pixel would have been visible in the previous views, near  the other background pixels. The ``neighbor" in the lower level  is thus the pixel just next to the disoccluded area.
\end{itemize}
We describe now precisely how each of the pixel types in Fig.~\ref{fig:typesOfconnextions} is handled in our graph-based representation.   First, for the appearing pixels, their corresponding values, $\gamma^r_{2,j}$, are assigned without any connectivity information (they are implicitly attached to the side of the image). In the example of  Fig.~\ref{fig:intuitiveexample}, we see that the dark blue appearing pixel is stored in  level $2$ at its position in $I_{2}$, \emph{i.e.,} it corresponds to $\gamma^r_{2,1}$. 
Similarly, for the disoccluded pixels, since they do not appear in the reference image, their color value is stored in the position $\gamma^r_{2,j}$ in the color matrix, where $j$ corresponds to the pixel positions in the view $I_{2}$. In Fig.~\ref{fig:intuitiveexample}, the disoccluded pixels are stored in $\gamma^r_{2,3}$ and $\gamma^r_{2,4}$.
Additionally, at the reference level and at the position $c$ of the last pixel before the foreground object on row $r$, we store the connection value $\lambda^r_{1,c} = d+1$, where $d$ is the disparity vector associated to depth value $Z_1(r,c)$. This connection value links the last background pixel of the reference view to the ones in the target view that are disoccluded. For example, in Fig.~\ref{fig:intuitiveexample}, the foreground object is red. In level $1$, the last pixel before this foreground object is at position $1$ (light blue pixel). The graph thus links this pixel to the first disoccluded pixel of level $2$. These two pixels are considered as neighbors. The disparity $d$ of the background in the example of Fig.~\ref{fig:intuitiveexample} is equal to $1$, so the connection value is equal to $d+1 = 2$, \emph{i.e.}, $\lambda^r_{1,1} = 2$.

The occluded pixels correspond to a jump in the reference view, since they represent color values that are absent in the second view, and only visible in the reference one. The jump value is stored in the connectivity matrix at the following two positions: 1) the last pixel of the foreground object (with a connection value equal to the foreground disparity) and 2) the last pixel of the corresponding occluded region (with a connection value equal to the background disparity). In the example of Fig.~\ref{fig:intuitiveexample}, the last pixel of the foreground object 1) is at position $4$ in level $1$. Thus, we have $\lambda^r_{1,4} = 3$, since the red foreground object has a disparity of $3$. Secondly, the last pixel of the occluded region 2) is at position $6$ in level $1$. Since the background disparity is $1$, we have $\lambda^r_{1,6} = 1$. We notice that the two connections meet at $c=7$ in level $2$, which corresponds to the position of the last foreground pixel in level $2$. This time, since no new pixel is contained in the second view, we do not store any value in the color vector. 
Finally, the disappearing pixels are simply indicated by a connection value at the position of the last preceding pixel. This connection value is equal to the background disparity. In the example of Fig.~\ref{fig:intuitiveexample}, the first disappearing pixel is at position $19$, thus $\lambda^r_{1,18} = 1$. For the next views, the graph construction proceeds in the same way, \emph{i.e.}, each view is connected to the previous one and constitutes a new level in the graph. This leads to $H$ matrices $\Lambda_r$ and $\Gamma_r$ ($H$ is the number of rows) that are concatenated in two 3D matrices $\Lambda$ and $\Gamma$ and constitute the complete GBR data structure.

The GBR construction strategy introduced above is presented in a general form in Algorithm~\ref{alg:gbr}. The inputs are two luminance views $I_1$ and $I_2$, the depth image $Z_1$ and the distance between the two cameras $\delta$. First, we convert the depth image  into a dense disparity map (line \ref{lin:DispConvertion} to \ref{lin:DispConvertionEnd}). The non-integer disparity value is simply rounded to the closest integer since the current GBR implementation only handles such values. This operation induces an approximation error that is corrected by a residual image as detailed in Section~\ref{sec:coding}. Then, the graph construction is done row by row. The pixels of $I_1$ are first inserted in the first level of the luminance matrix (lines \ref{lin:FirstLevel} to \ref{lin:FirstLevelEnd}). We then insert the appearing pixels on level $2$ of the luminance matrix (lines \ref{lin:appearing} to \ref{lin:appearingEnd}). After this operation, we go through the dense disparity map of $I_1$ and detect disocclusions (lines \ref{lin:disocclusions} to \ref{lin:occlusions}) and occlusions (lines \ref{lin:occlusions} to \ref{lin:occlusionsEnd}). For building a graph with more than $2$ images, one simply needs to repeat the operations from lines \ref{lin:appearing} to \ref{lin:levelEnd} for every predicted view, while taking as starting point the most recent view. Finally, the matrices for every row are concatenated in the 3D matrices $\Lambda$ and $\Gamma$.

With the above graph construction method, the graph representation is sparse (only a small fraction of entries is non-zero) and avoids all redundancy in the color value description since the pixels values stored at a given level in $\Gamma_r$ are only those that are not present in the lower levels. Another important advantage of this graph representation consists in the multi-level structure, where the connections in one level are related to connections in other lower levels and for a chain of connections. Therefore, a reconstruction algorithm  only needs to  go through these connection chains to reconstruct the different multiview images.

\begin{algorithm}
\singlespacing
\footnotesize
\caption{GBR construction for two levels}\label{alg:gbr}
\begin{algorithmic}[1]
\REQUIRE{
\STATE $\{I_1, I_2\}$ -   luminance images of height $H$ and width $W$\;
\STATE     $Z_1$ -  the depth map corresponding to view $1$\;
\STATE     $\delta$ -  the distance between the two views\;
}
\ENSURE{The color and geometry matrices $\Gamma$ and $\Lambda$\;}\\
 \item[] \hspace*{-1.8\baselineskip} \textbf{Algorithm:}
\item[]  \COMMENT{Convert depth $Z_1$ to dense disparity map $D$ with rounding operation}
\FOR{$ r \gets 1 \textrm{ to } H$ and $ c \gets 1 \textrm{ to } W$} \label{lin:DispConvertion}
\STATE $ D(r,c) \gets \ceil[\Big]{ \frac{f\delta}{Z(r,c)} + 0.5} $ \; \label{lin:rounding} 
\ENDFOR \label{lin:DispConvertionEnd}
\item[]
\FOR{$ r \gets 1 \textrm{ to } H$} 
\item[]
\item[]  \COMMENT{Insert $I_1$ in the first level of the color matrix $\Gamma_r$}
\FOR{$ c \gets 1 \textrm{ to } W$} \label{lin:FirstLevel}
\STATE $ \Gamma_r(c,1) \gets I_1(r,c) $ \; 
\ENDFOR \label{lin:FirstLevelEnd}
\item[]
\item[] \COMMENT{Insert the $D(r,1)$ appearing pixels ((a) in Fig.~\ref{fig:intuitiveexample}) in the second level of $\Gamma_r$}
\FOR{$ c \gets 1 \textrm{ to } D(r,1)$} \label{lin:appearing}
\STATE $\Gamma_r(c,2) \gets I_2(r,c)$ 
\ENDFOR \label{lin:appearingEnd}
\item[]
\STATE $ c_1 \gets 2$ \COMMENT{current column index in $I_1$}
\STATE $ d_p \gets D(r,1) $ \COMMENT{previous disparity value}
\STATE $c_{\rm stop} \gets D(r,c_1)+1$ \COMMENT{column index in level $2$ that serves as stopping criterion}
\item[]
\WHILE{$c_{\rm stop} \leq W$} 
\STATE $d_c \gets D(r,c_1)$ \COMMENT{current disparity value, $\neq d_p$ in the case of occlusion or disocclusion}
\IF{$d_c \neq d_p$}
		\STATE $\Delta_{\rm disp} = d_c-d_p$ \COMMENT{disocclusion ($>0$) or occlusion  ($<0$) size}
		\IF{$\Delta_{\rm disp}>0$}\label{lin:disocclusions}
			\STATE $c_{\rm stop} \gets c_{\rm stop}+ \Delta_{\rm disp}$
			\item[] \COMMENT{Fill the disoccluded pixels ((b) in Fig.~\ref{fig:intuitiveexample}) in the second level of  $\Gamma$ }
			\FOR{$ c_2 \gets c_1+ d_p \textrm{ to }  \min(c_1+ d_p+\Delta_{\rm disp}-1,W)$}
				\STATE $\Gamma(r,c_2,2) \gets I_2(r,c_2)$
			\ENDFOR
			\item[] \COMMENT{Include the link between the two neighbors in the 3D space ((b) in Fig.~\ref{fig:intuitiveexample}) in $\Lambda$ }
			\STATE $\Lambda(r,c_1-1,1) \gets d_p+1$
		\ELSE \label{lin:occlusions} 
		 	\item[] \COMMENT{Include the jump ((c) in Fig.~\ref{fig:intuitiveexample}) in $\Lambda$}
			\STATE  $\Lambda(r,c_1-1,1) \gets d_p$
			\STATE  $\Lambda(r,c_1-1+|\Delta_{\rm disp}|,1) \gets d_c$
			\STATE $c_1 \gets c_1+|\Delta_{\rm disp}|$
		\ENDIF \label{lin:occlusionsEnd}
\ELSE
	\STATE $c_{\rm stop} \gets c_{\rm stop} +1$
\ENDIF

\STATE $ d_p \gets d_c $
\STATE $c_1 \gets c_1+1$
\ENDWHILE
\item[]
\STATE  $\Lambda(r,c_1-1,1) \gets W - c_1 + 1$  \COMMENT{disappearing pixels ((d) in Fig.~\ref{fig:intuitiveexample}) }
\ENDFOR \label{lin:levelEnd}
\item[]
\FOR{$r \gets 1 \textrm{ to } H$ and  $c \gets 1 \textrm{ to } H$ and $l \gets 1 \textrm{ to } 2$}
\STATE $\Gamma(r,c,l) \gets \Gamma_r(c,l)$
\STATE $\Lambda(r,c,l) \gets \Lambda_r(c,l)$
\ENDFOR
\end{algorithmic}
\end{algorithm}

\subsection{View reconstruction at the decoder}
The graph information described in the previous section is used directly for view reconstruction at the decoder, which has access to graph components $\Gamma_r$ and $\Lambda_r$ for every row $r$. The reconstruction of a certain view requires the color values and the connections of all lower levels. The reconstruction of the color values in the current view is performed by navigating the graph between the different levels. This navigation starts from the border of the image at the level that needs to be constructed, then follows the connections and refers to the lower levels when no color information is available at current level.
We show in Fig.~\ref{fig:reconstructionexample} an example of a view synthesis for the image of level $2$, based on the graph in the example of Fig.~\ref{fig:intuitiveexample}. The pixel numbering is done with respect to the column index of $I_2$, as in Fig.~\ref{fig:intuitiveexample}. The reconstruction starts with the appearing pixel $1$ at level $2$ . Then, it moves to the reference level and fills pixel color values until encountering a non-zero connection. The first connection is after pixel $2$ and links it to pixel $3$ and $4$ in level $2$. After filling all the disoccluded pixels, the reconstruction goes back to the reference level and fills color information ($5$, $6$ and $7$) until the next non-zero connection (at pixel $7$). The connection in $7$ indicates an occluded region. Hence, the reconstruction algorithms jumps across columns in the reference view and continues the decoding of the pixels in the reference level for pixel $8$ to $19$ until it encounters the next non-zero connection (disappearing pixel). The reconstruction of the other views (\emph{i.e.}, the other levels of the graph) is done recursively.
We see that the reconstruction process is very simple and that the required geometry information is captured in a flexible and controlled way by the graph connections. The integer disparities obtained after a rounding operation leads however to errors in the view prediction. We leave for future work the study of more evolved techniques that could interpolate pixels from float disparities, as it is done in \cite{Mao_Y_2013_picip_exp_hfdibrgbr}. In this paper, we handle these rounding errors with the generation of residual images, as detailed in Section~\ref{sec:coding}.

\section{GBR information coding}\label{sec:coding}

In this section, we propose a complete encoding scheme where the GBR information (color and geometry) is compressed to provide a compact description of multiview images. The geometrical errors, due to depth errors or non-integer disparities, are carefully taken into account in order to minimize the reconstructed image distortion.

\subsection{Geometry coding}

We describe first our approach to code the graph connections. As we can observe in the example of Fig.~\ref{fig:intuitiveexample}, the matrix of connections for row $r$, $\Lambda_r$, is sparse. Hence, it can be coded with a small number of bits. For that purpose, we do not code directly the connection matrix $\Lambda_r$ and rather consider a small matrix $\Phi$ of size $M\times 4$, where $M$ is the number of non-zero elements in all the the connection sub-matrices $\Lambda_r$ with $r<H$. The matrix $\Phi$ stores all the meaningful connections which are characterized by $4$ parameters. The first column of $\Phi$ contains the row indices $r$ for each graph connection. The second column contains the  column indices $c$, the third column contains the graph level indices, and finally, the fourth column contains the actual connection values. We then code the columns separately using first a differential operator along each of them in order to decrease the entropy, and then an arithmetic coding technique.

\begin{figure}[t]
\centering
\begin{minipage}{0.24\linewidth}
\center
		\includegraphics[width=1\linewidth]{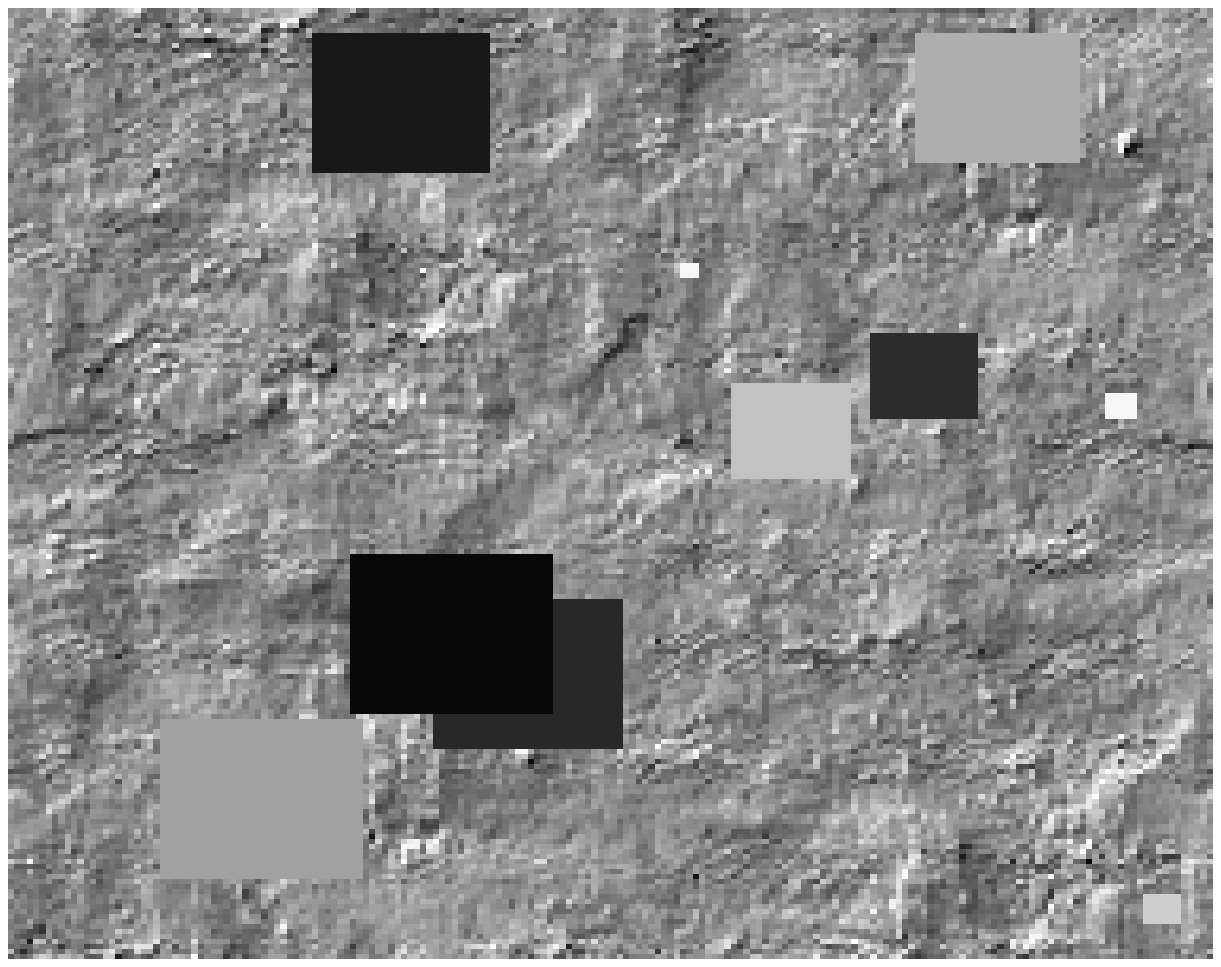}
\end{minipage}\hfill
\begin{minipage}{0.24\linewidth}
\center
		\includegraphics[width=1\linewidth]{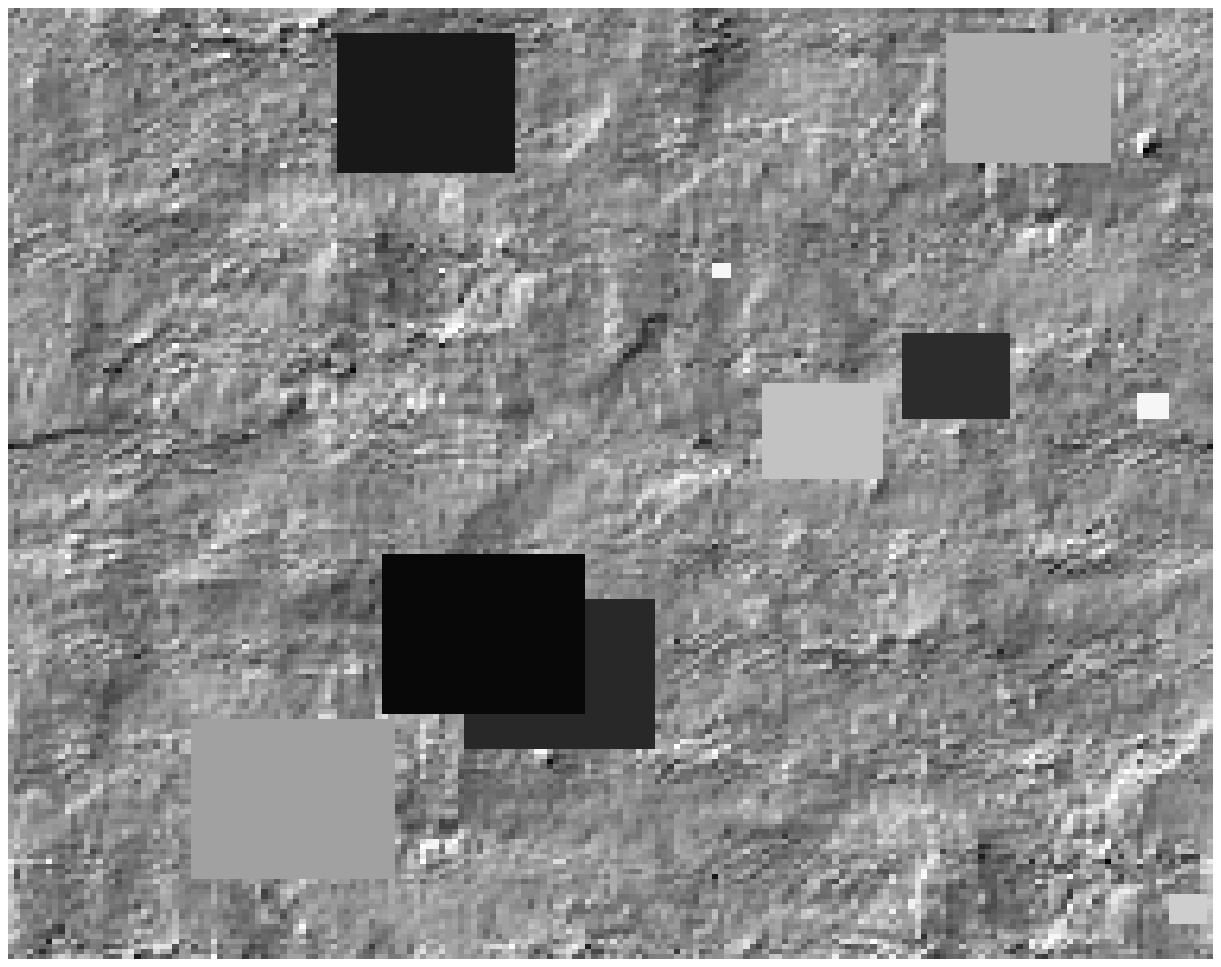}
\end{minipage}\hfill
\begin{minipage}{0.24\linewidth}
\center
		\includegraphics[width=1\linewidth]{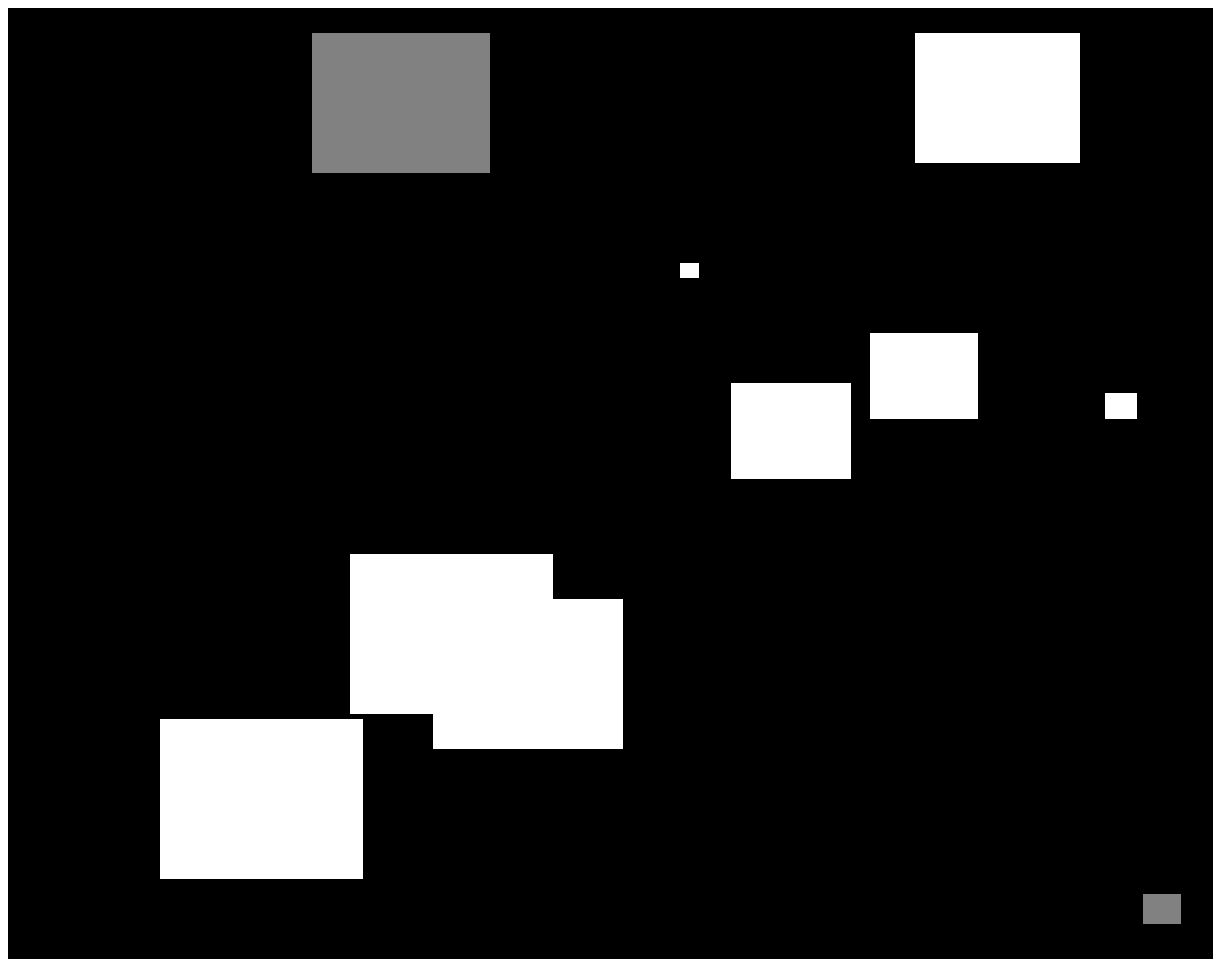}
\end{minipage}\hfill
\begin{minipage}{0.24\linewidth}
\center
		\includegraphics[width=1\linewidth]{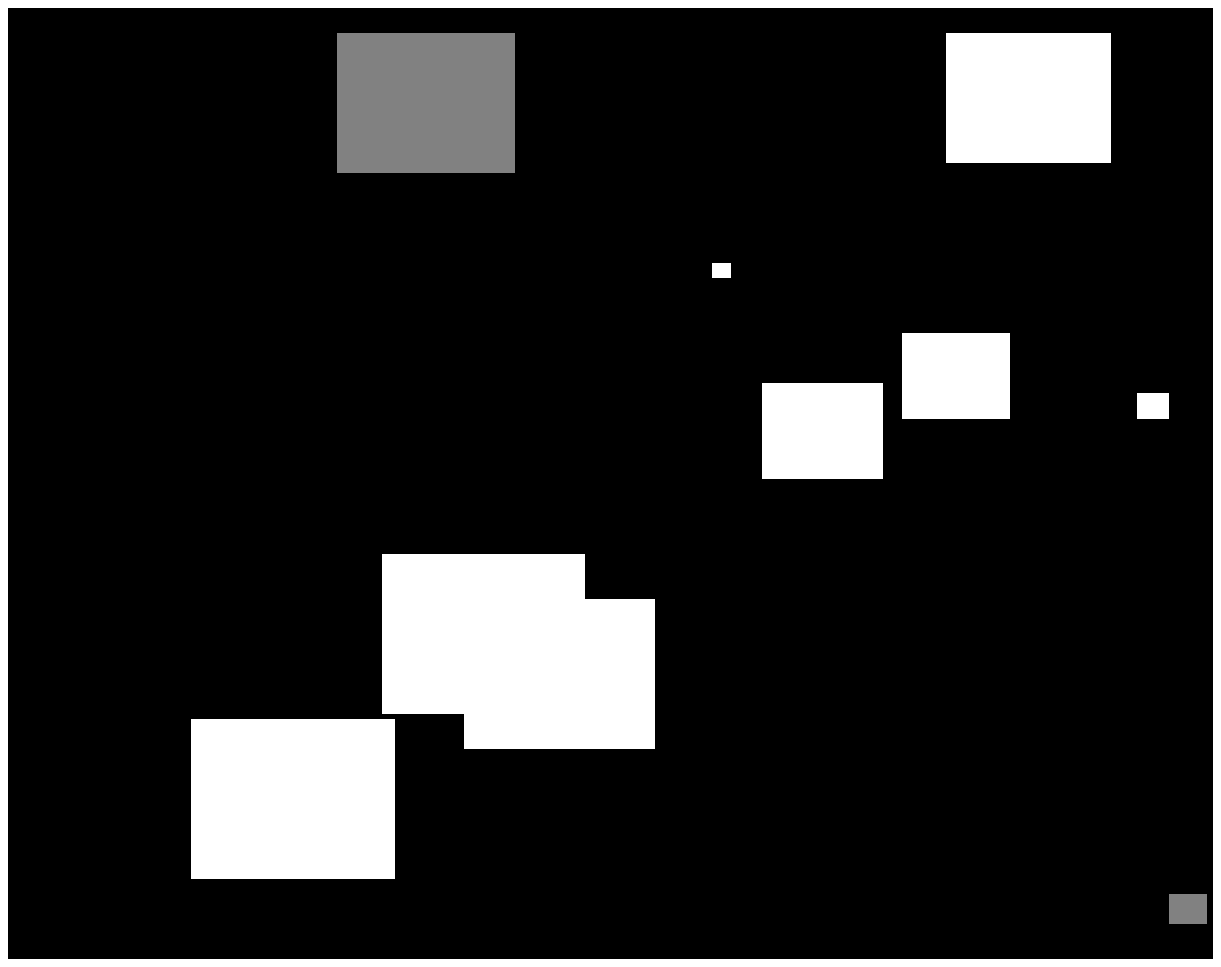}
\end{minipage}\hfill
\begin{minipage}{0.24\linewidth}
		\centerline{view 1 (luminance)}
\end{minipage}\hfill
\begin{minipage}{0.24\linewidth}
		\centerline{view 2 (luminance)}
\end{minipage}
\begin{minipage}{0.24\linewidth}
		\centerline{view 1 (depth)}
\end{minipage}\hfill
\begin{minipage}{0.24\linewidth}
		\centerline{view 2 (depth)}
\end{minipage}
	 \caption{View $1$ and $2$ luminance and disparity  images of the ``squares 1" dataset. The disparities of the objects in the scene are integer.}
	\label{fig:exp1}	
\end{figure}

\begin{figure}[t]
\centering
		\includegraphics[width=0.5\linewidth]{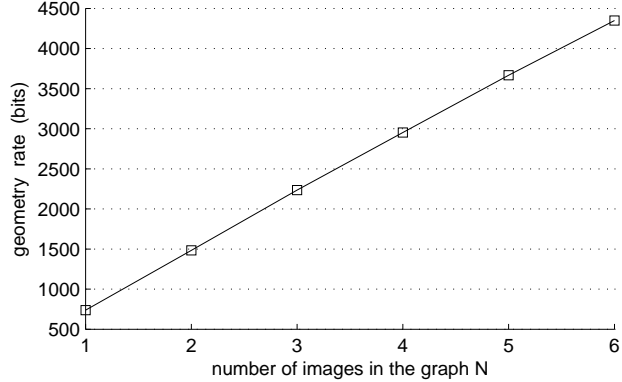}
	\caption{Evolution of the coding size (in bits) for lossless geometry compression, as a function of the number of images $N$ in the ``squares 1" dataset.}
	\label{fig:gbrVsLevel}
\end{figure}

We illustrate the behavior of the proposed compression scheme in the lossless coding of images of an artificial dataset (called ``squares 1", $190\times190$, and shown in Fig.~\ref{fig:exp1}). The data represents a 3D scene made of one plane background and multiple  foreground square objects. The scene is captured by $N$ parallel cameras such that the disparities of the objects between the viewpoints is only horizontal. The background and the foreground objects are parallel to the camera planes. The number of foreground objects, their position, their size and the integer disparity values are generated randomly.  We show in Fig.~\ref{fig:gbrVsLevel}, the evolution of the graph geometry coding size (bits) as a function of the number of views $N$ involved in the representation. 
Although the observed linear relationship only depends on the regularity of the scene and acquisition, we notice that the required number of bits increases with the number of levels. This is due to the nature of the graph construction. It reflects that GBR sends ``just enough" geometry information for a given number of views to predict, and increases this geometry precision as soon as it becomes higher.

In order to decrease the coding costs, we can estimate the geometry in some views, instead of coding it for every image. Hence, we introduce the possibility of removing some images (\emph{i.e.}, levels) from the graph structure and interpolating them at the receiver. In this case, fewer bits are required for encoding the geometry since the number of levels is reduced. When a level is removed from the graph, the graph links are directly extended to the next level (\emph{e.g.}, edges connect levels $2$ and $4$ directly, instead of passing through level $3$), and the pixel values of the level that is skipped are stored in the upper level. 
However, the interpolation of views at the decoder may create some distortion in the geometry.  The interpolation of a view at the decoder  is done by disparity compensation with the two closest received images. The two disparity-compensated estimations of the interpolated view are then merged, which results in a synthesized image with no disocclusion. Since the disparity maps are not explicitly transmitted in a GBR-based scheme, they are retrieved from the values of the connections in the graph. In other words, the GBR  geometry can be used for virtual view synthesis at the decoder, similarly to what can be achieved with depth images. The choice of the number of levels $L$ and of which levels are included in the graph is a tradeoff between the bitrate required for graph transmission, and the distortion of the reconstructed view distortion induced by view removal. In this paper, we choose $L$ and the views with a full search algorithm that evaluates the graph size and the rendered distortion for many configurations.

\subsection{Luminance compression and residual images}

The color signal compression may benefit from the graph structure that links pixels at different levels to each other. In the proposed scheme, the reference image is encoded with traditional image coding tools. The novel pixels at every level are to be coded by traversing pixels along the graph connections. One of the interesting properties of the graph is that it links pixels that are supposed to represent two neighboring points in the 3D scene. In other words, these pixels might be correlated, which can be exploited for coding. For the current system, we use a simple differential operator along the graph. The differentiated color values are then coded using an arithmetic coder. Development of more sophisticated graph-based techniques is part of our ongoing work.

\begin{figure}[t]
\centering
		\includegraphics[width=0.6\linewidth]{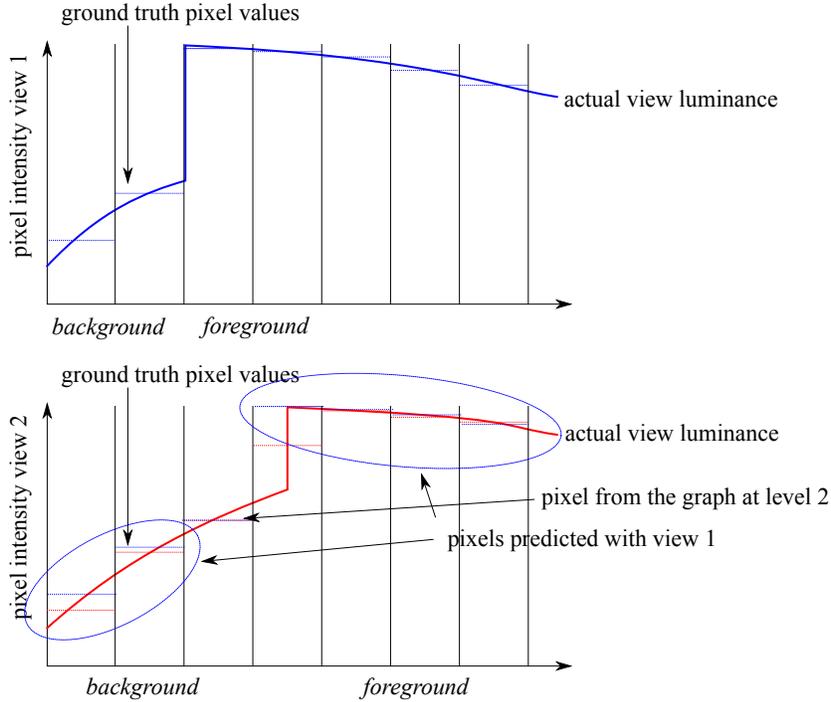}
	\caption{Illustration of prediction error between two views when half-pixel disparity occur for a given row $r$.}
	\label{fig:prediction}
\end{figure}

\begin{figure*}[t]
\centering
\begin{minipage}{0.32\linewidth}
		\includegraphics[width=1\linewidth]{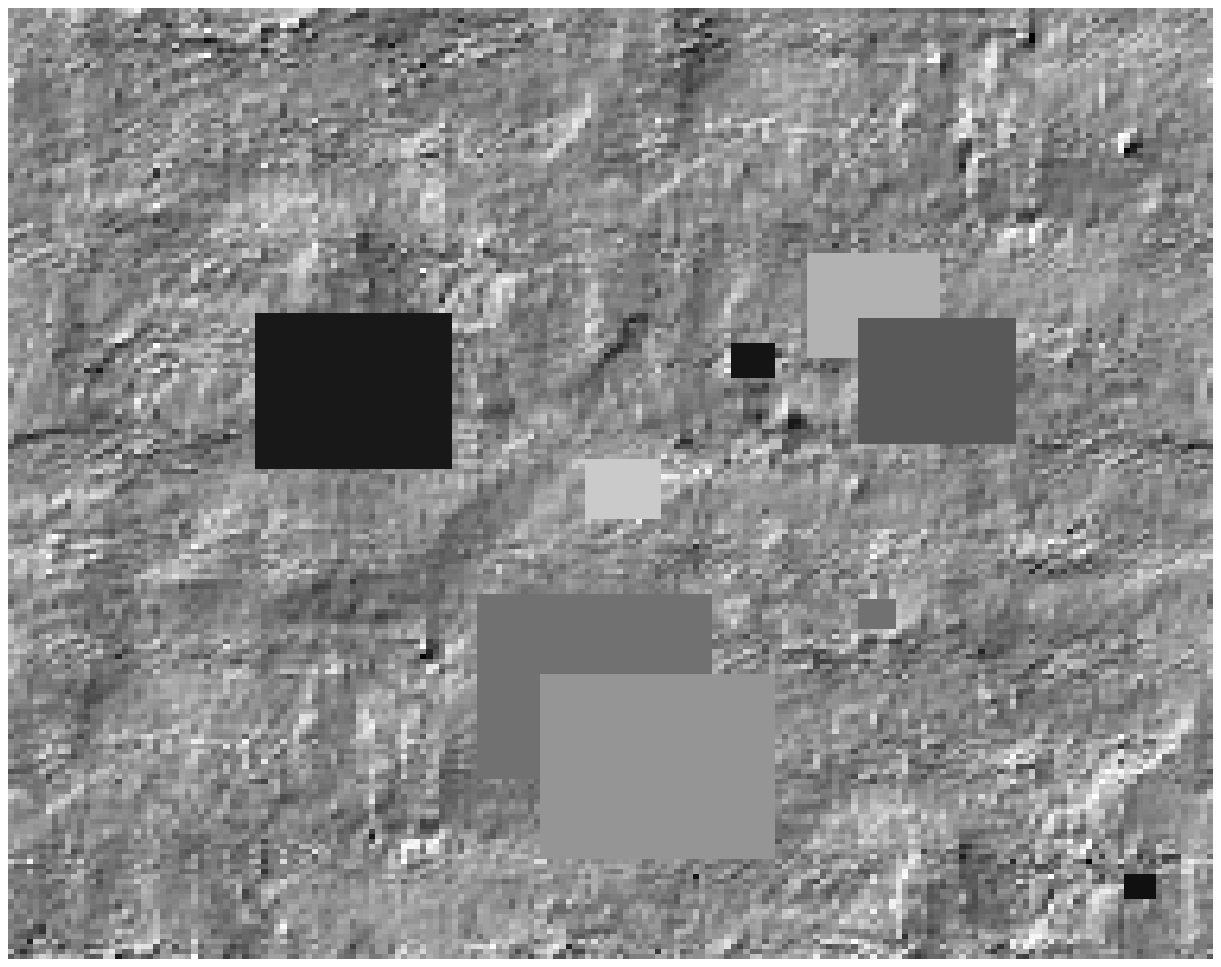}
		\centerline{(a)}
\end{minipage}\hfill
\begin{minipage}{0.32\linewidth}
		\includegraphics[width=1\linewidth]{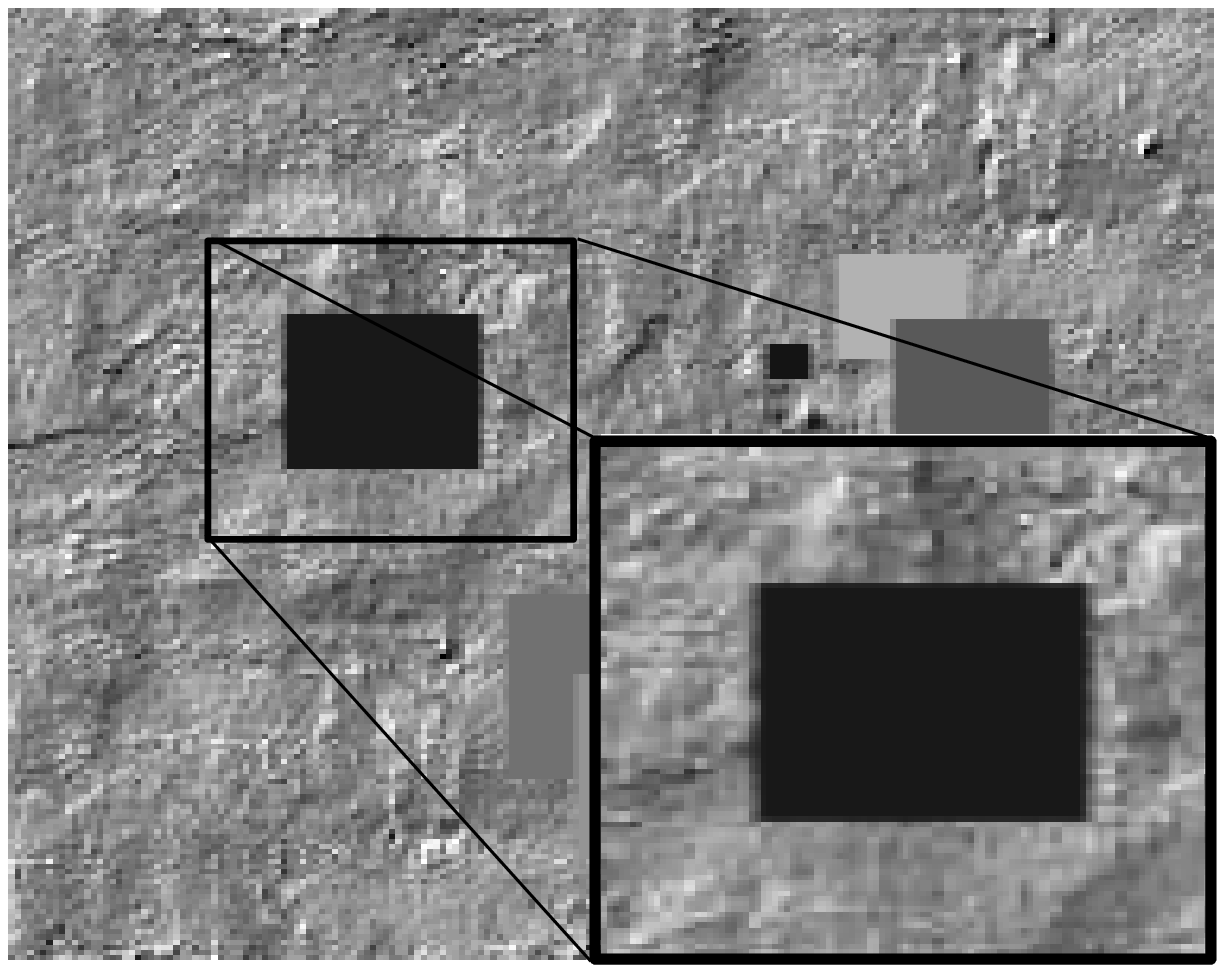}
		\centerline{(b)}
		\end{minipage}
\begin{minipage}{0.32\linewidth}
		\includegraphics[width=1\linewidth]{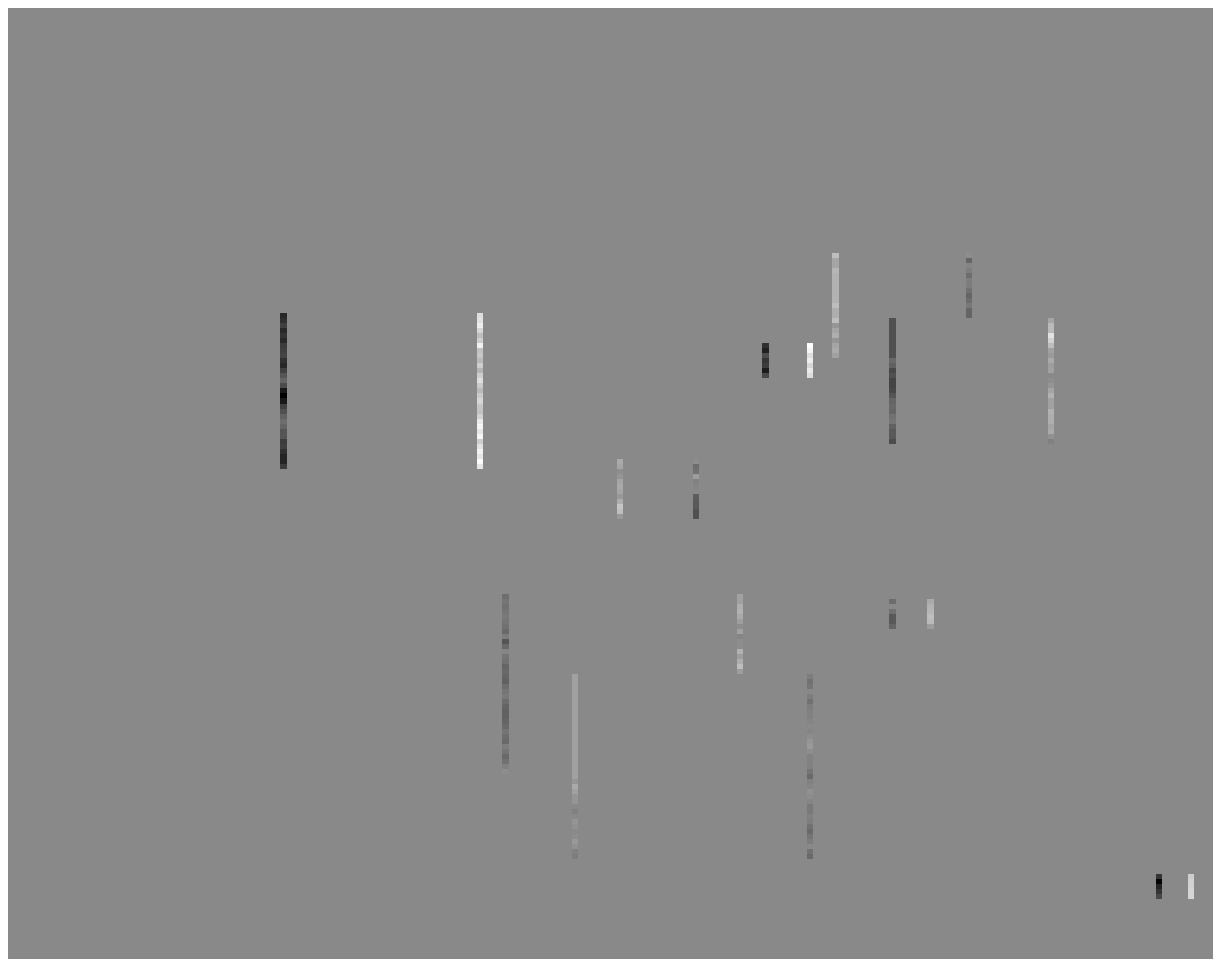}
		\centerline{(c)}
\end{minipage}
	\caption{Views $1$ (a) and $2$ (b) and residual error image of view $2$ (c) after geometry prediction from GBR connections for the ``squares 2" dataset. The zoomed foreground object is shown for view $2$ (b). The disparities of the objects in the scene are non integer, such that error appear at object boundaries as it can be observed in (c).}
	\label{fig:exp2}	
\end{figure*}

The graph  introduced in the previous sections only handles integer disparities in the connections. However, the actual disparity value obtained from depth data might not always be integer and the rounding operation (see line \ref{lin:rounding} of Algorithm~\ref{alg:gbr}) in the graph construction brings a geometric error in the view prediction. In addition, if the initial depth map contains errors, the  views predicted by geometric projection contain error too. In order to compensate for these errors, we generate residual images that correct the view prediction error and prevent error propagation through the graph levels.  
Compensation errors may appear on almost every part of the predicted images as illustrated in Fig.~\ref{fig:prediction}. More precisely, sub-pixel precision disparity values  imply a pixel corresponding to an object in the scene is not necessarily captured at integer pixel positions by two adjacent viewpoints. Therefore, when a view is predicted from another one, errors may occur. We can see in Fig.~\ref{fig:prediction} that these errors may appear in every region of the image except in ones filled by the current level of the graph. However, we still build the residual at an image level with a value of zero at the location where the current level provides a new value.
 These residuals are coded using the JPEG2000 coder in our current implementation. In order to illustrate the role of the residual images, we build a dataset called ``squares 2" ($190\times190$) involving non integer disparities. The scene is made of square foreground objects with half pixel precision disparity values. As for the ``square 1" dataset, the position of the foregrounds and their disparity are initialized randomly. Thus for some views a foreground object may appear at half pixel position. In this case, the pixel intensity represented in the image is the average of the foreground and background luminance values. We show in Fig.~\ref{fig:exp2} the images corresponding to view $I_1$ and $I_2$ in the dataset, along with the residual error image of view 2, which is the difference between  $I_2$ and its estimation from $I_1$ and the given GBR geometry information. We see that the residual error image mostly contains energy at the object boundaries, as it is also the case in the illustration of Fig.~\ref{fig:prediction}.

Similarly, we also generate residual images to correct the error due to interpolation when a view is removed from the graph structure.  For the interpolated views, errors may occur in any region of the image, given that the connections of the graph do not correspond exactly to the actual geometry of the scene. Therefore, these residuals are again images, coded with JPEG2000.

We have described above a complete coding scheme where we can vary the coding precision of the color signal and of the residual images, and where we can also adjust the number of levels involved in the graph representation, in order to optimize the rate-distortion performance. We are thus able to generate several rate-distortion points from low to high bitrate. The optimal rate-allocation between the different components is however a complex task. 
In our prototype encoder, it relies on a full search algorithm between the different compression steps of all the components. Development of coding tools better suited for these datasets, as well RD optimization techniques, are part of ongoing work.

\begin{figure}[t]
\centering
\begin{minipage}{0.33\linewidth}
		\includegraphics[width=1\linewidth]{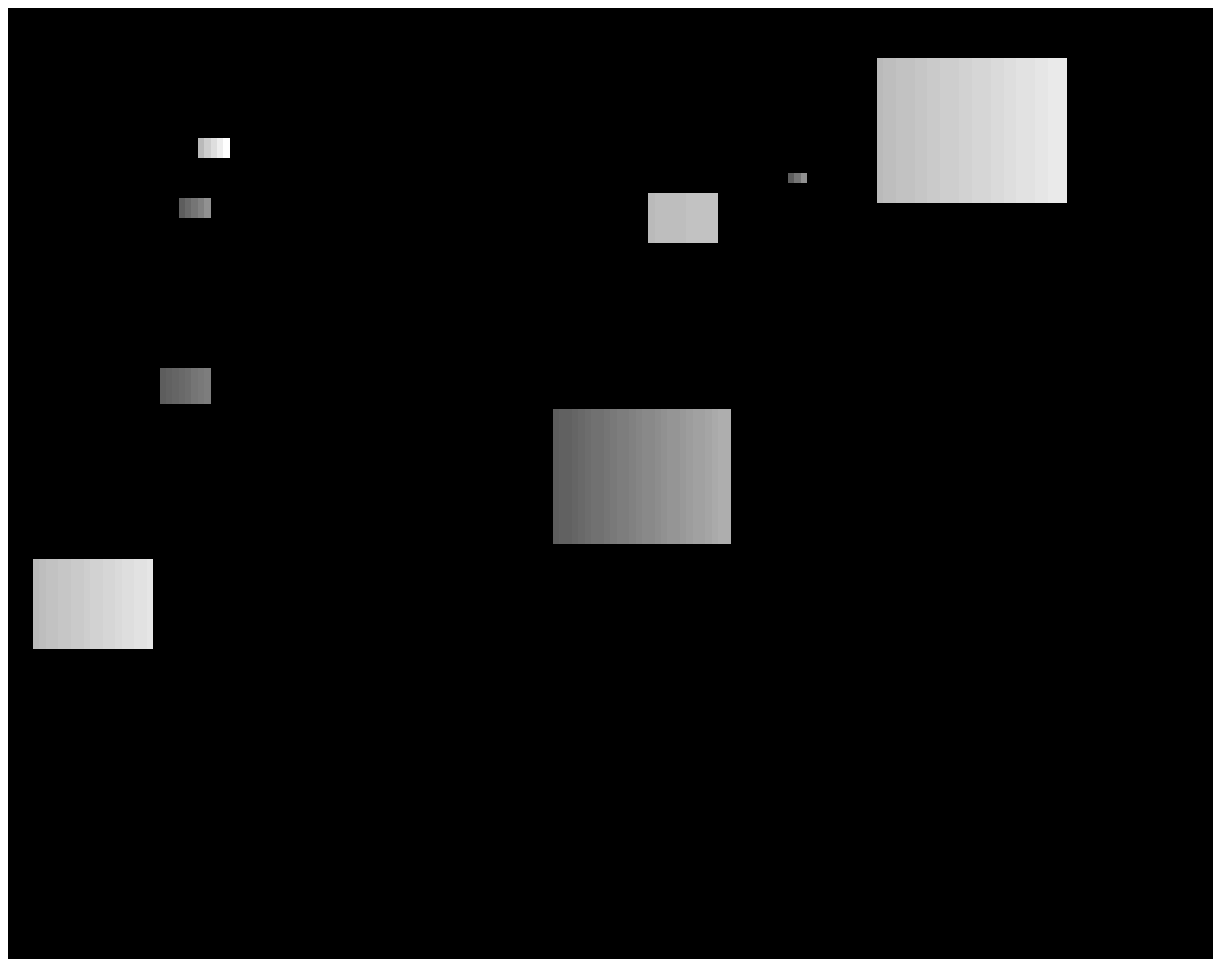}
		\centerline{(a) original depth map}
\end{minipage}\hfill
\begin{minipage}{0.33\linewidth}
		\includegraphics[width=1\linewidth]{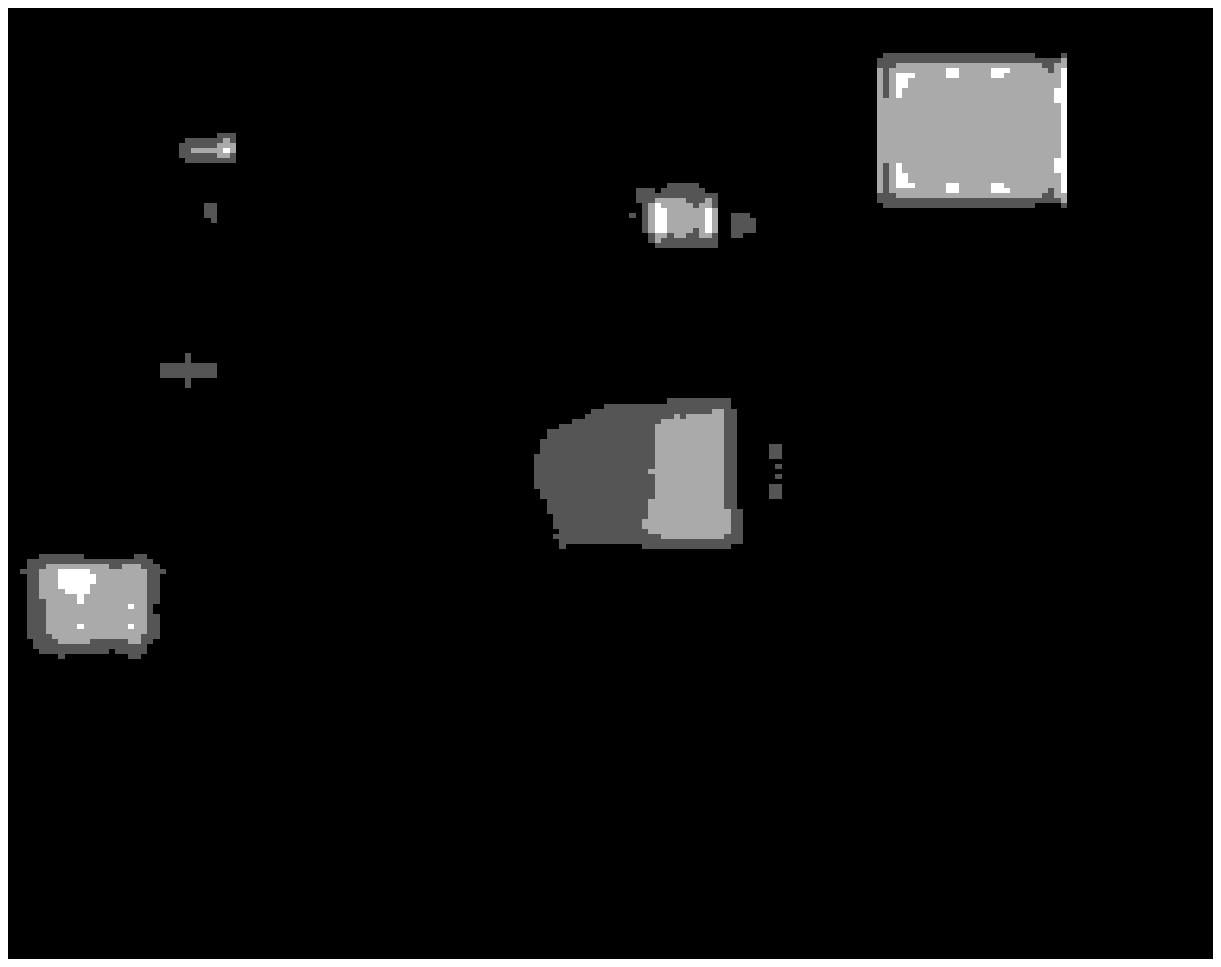}
				\centerline{(c) JPEG2000 }
\centerline{compressed depth map}
\end{minipage}\hfill
\begin{minipage}{0.33\linewidth}
		\includegraphics[width=1\linewidth]{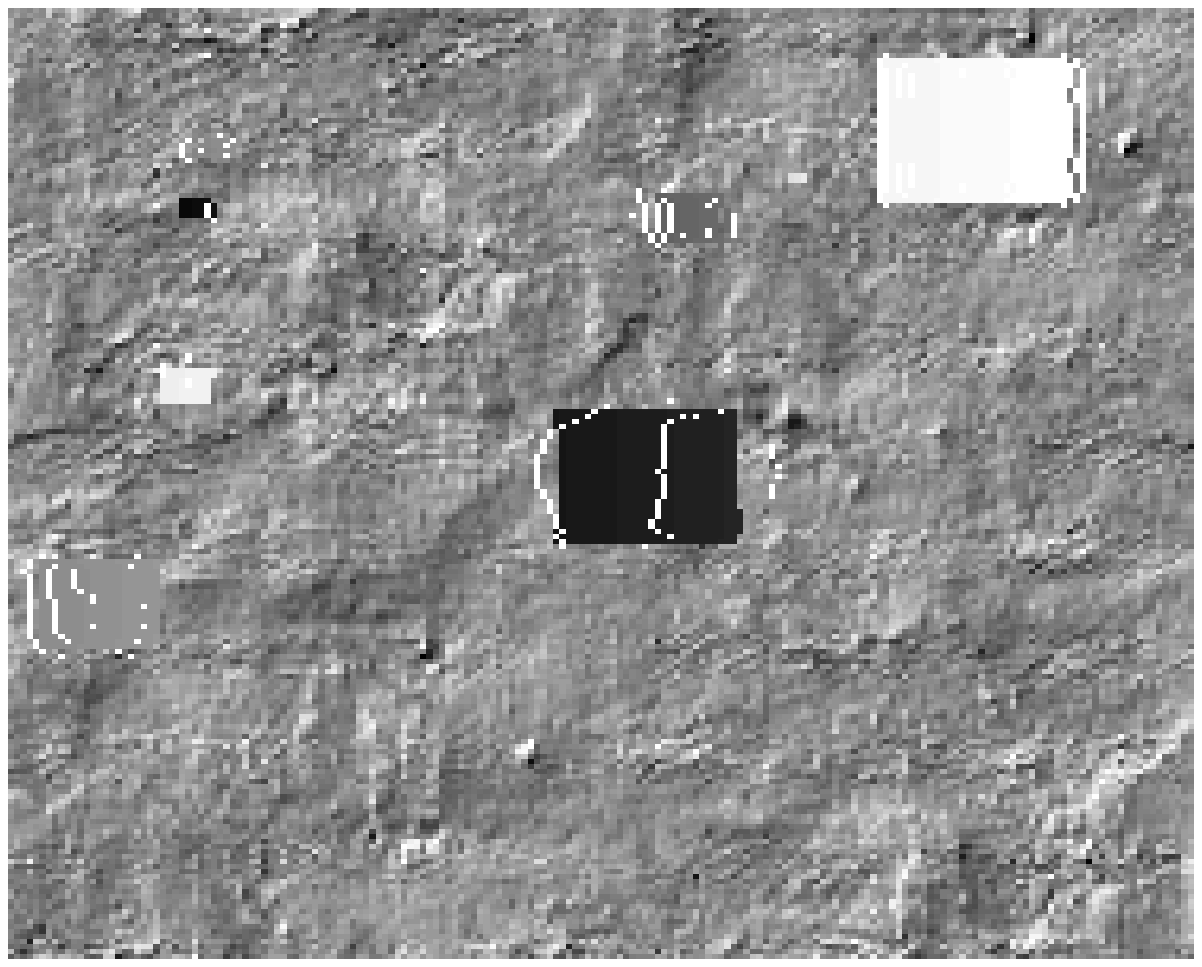}
				\centerline{(e) Reconstructed $I_2$}
				\centerline{from compressed depth map}
\end{minipage}\hfill
\begin{minipage}{0.33\linewidth}
		\includegraphics[width=1\linewidth]{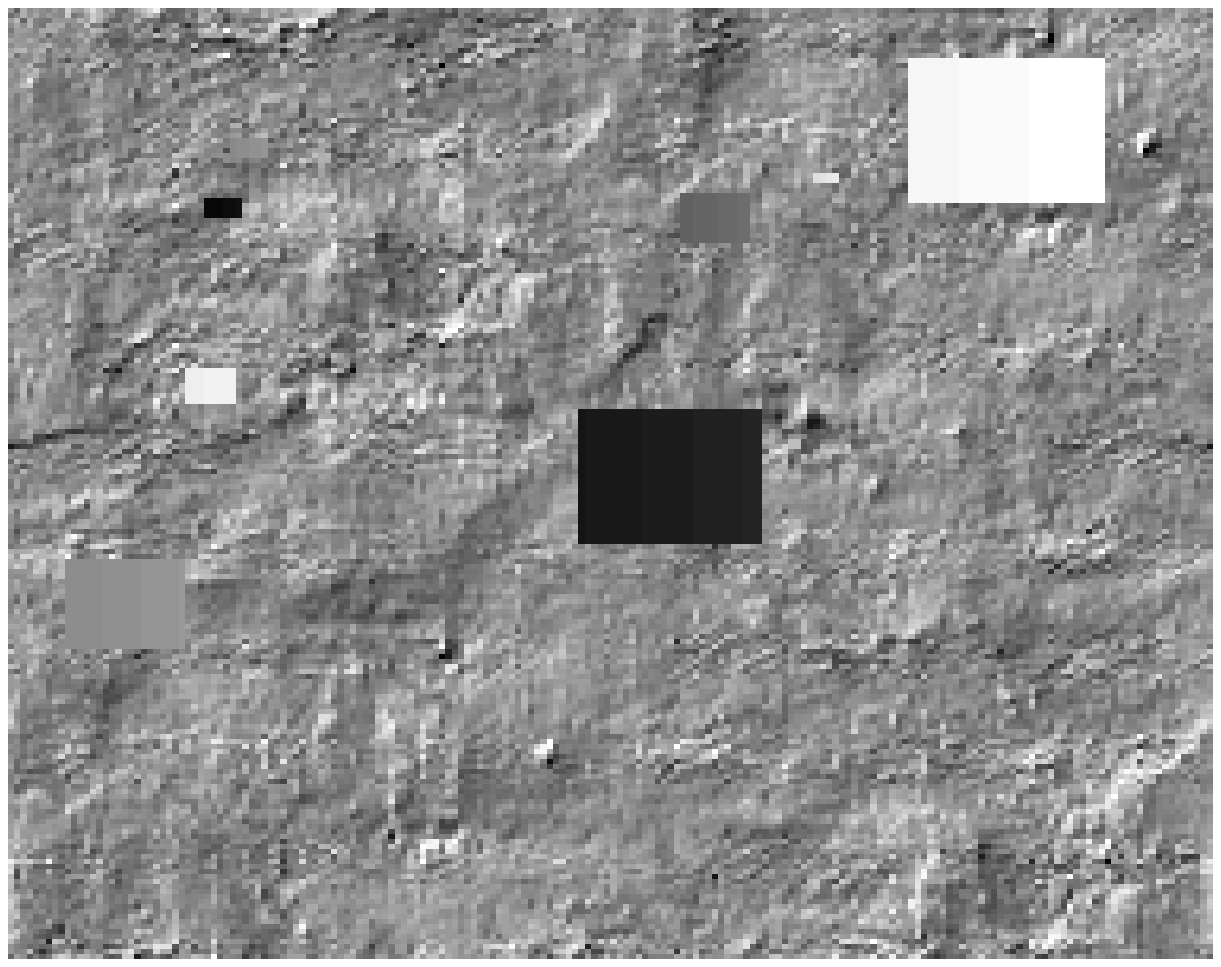}
		\centerline{(b) original $I_2$}
\end{minipage}
\begin{minipage}{0.33\linewidth}
		\includegraphics[width=1\linewidth]{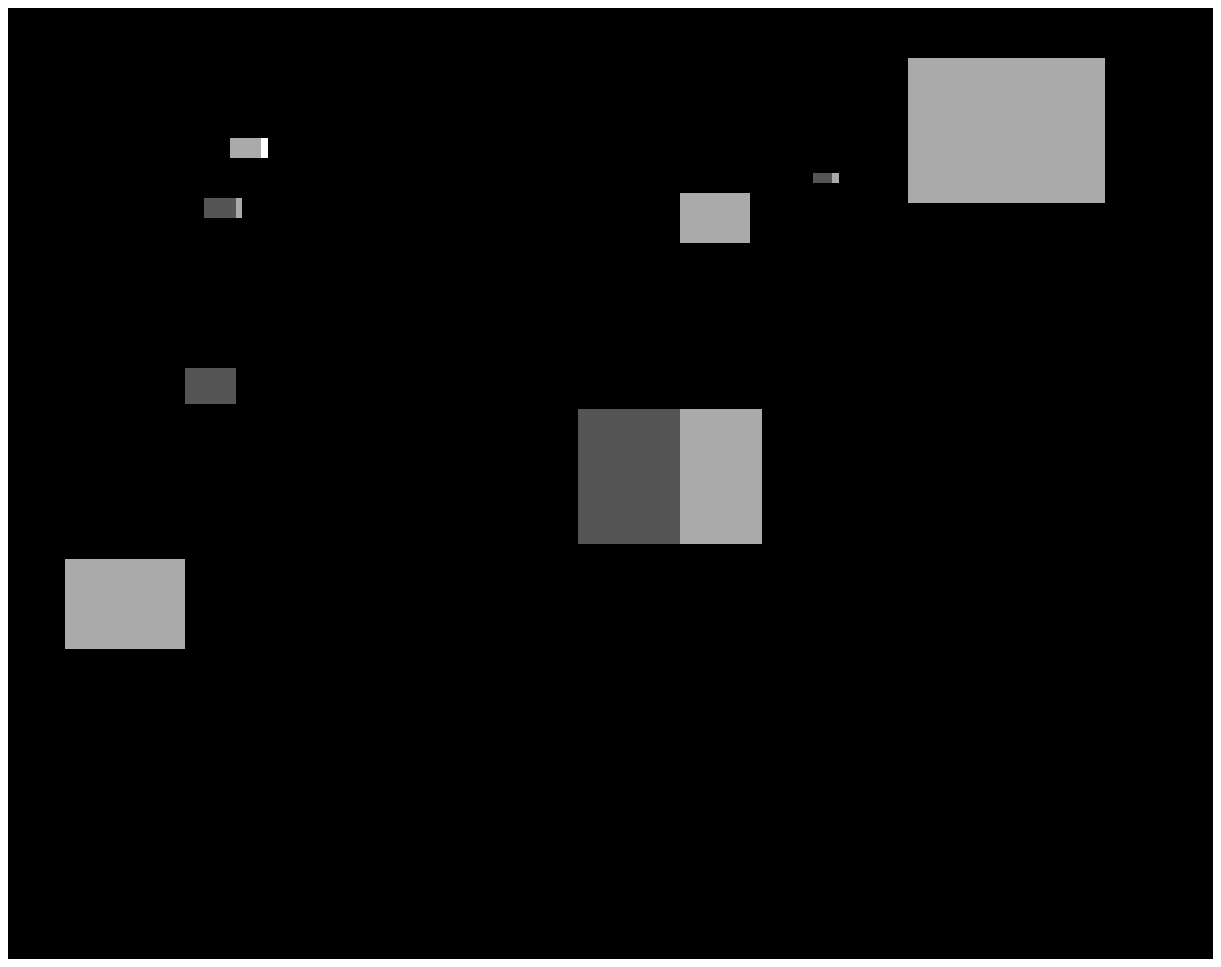}
				\centerline{(d) retrieved }
\centerline{disparity from GBR}
\end{minipage}\hfill
\begin{minipage}{0.33\linewidth}
		\includegraphics[width=1\linewidth]{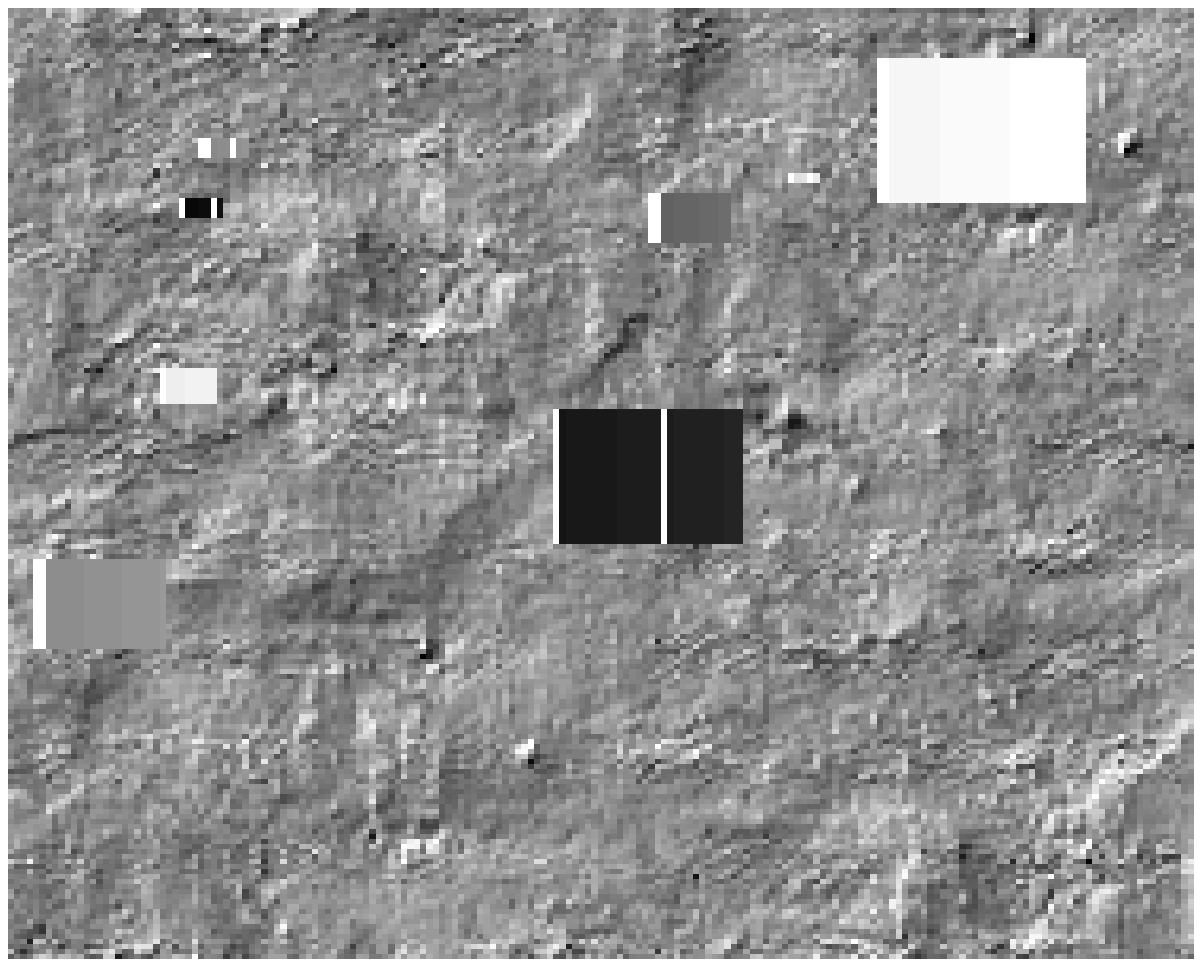}
				\centerline{(f) Reconstructed $I_2$ }
								\centerline{ from GBR geometry}

\end{minipage}
	\caption{Original depth of $I_1$ (a) and luminance of $I_2$ (b), JPEG2000 compressed depth map (c), and disparity map retrieved from GBR of the first view $I_1$ in the ``squares 3" dataset. Reconstruction of the second image $I_2$, from JPEG2000 compressed depth map (e) and GBR geometry (f). No residual error data is used for reconstruction.}
	\label{fig:exp3}	
\end{figure}

\begin{figure*}[t]
\centering
\begin{minipage}{0.32\linewidth}
		\center{\includegraphics[width=0.91\linewidth]{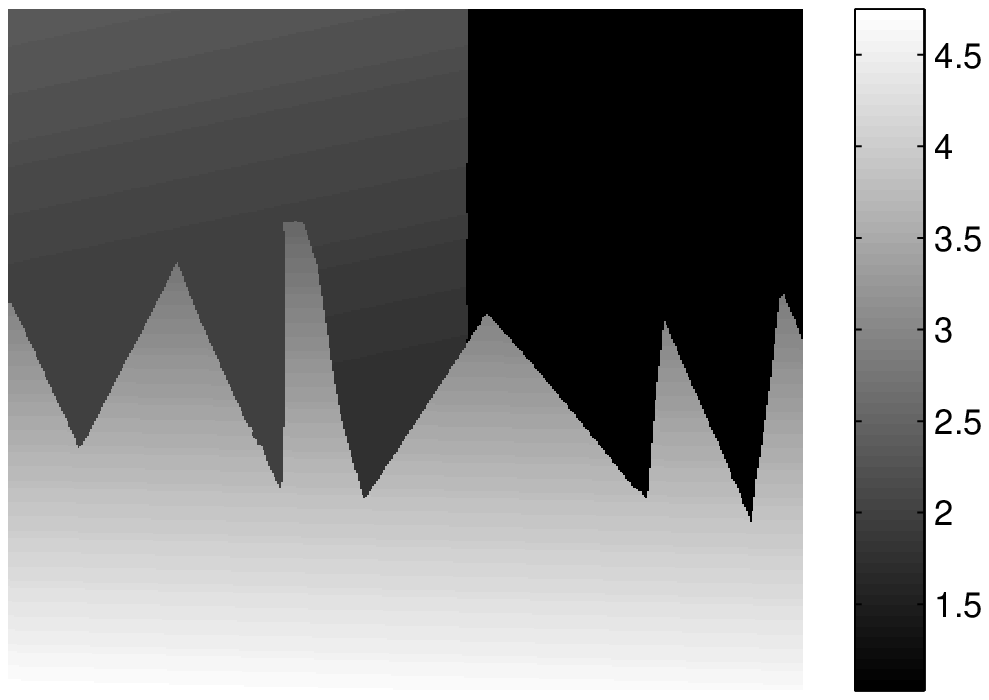}}
		\center{(a)}
				\end{minipage}
		\begin{minipage}{0.32\linewidth}
		\center{\includegraphics[width=0.91\linewidth]{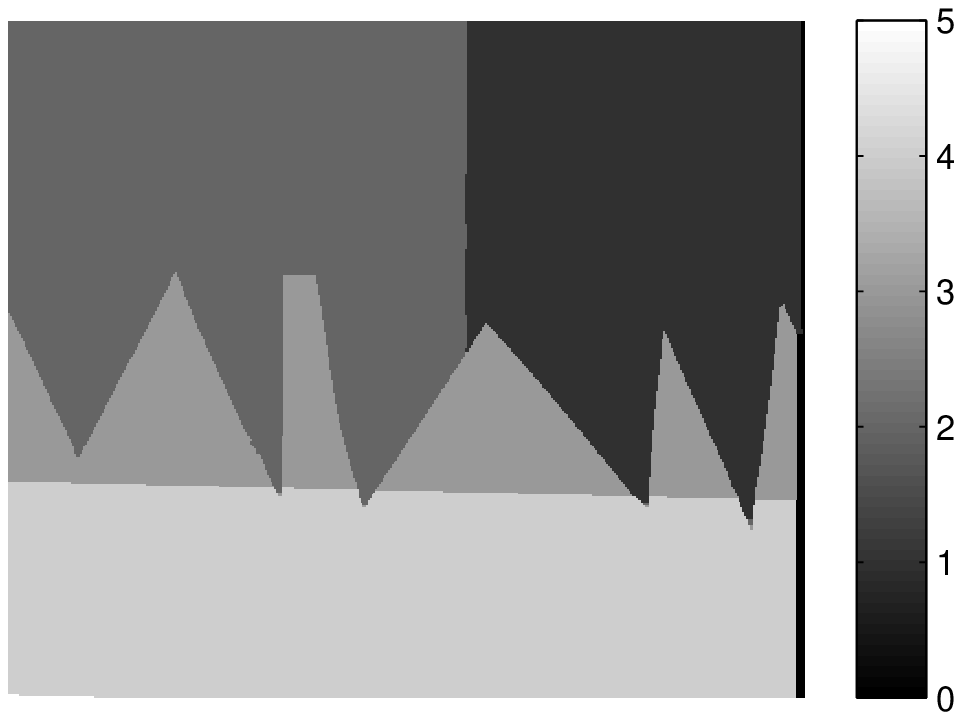}}
\center{(b)}
		\end{minipage}
		\begin{minipage}{0.32\linewidth}
		\center{\includegraphics[width=0.91\linewidth]{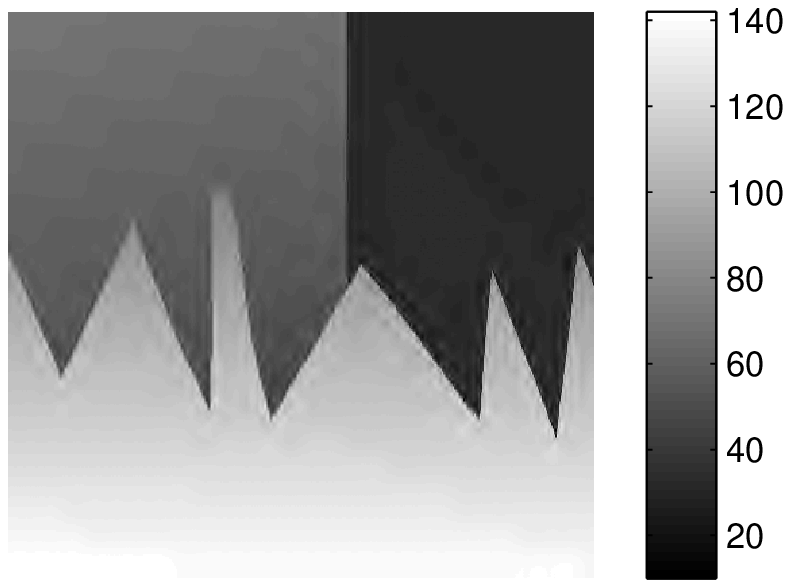}}
\center{(c)}
		\end{minipage}		
	\caption{Original depth image of ``sawtooth" dataset (a), the corresponding retrieved disparity from the GBR (b) and the depth image compressed with JPEG 2000 (c). GBR geometry and depth maps have been compressed at $0.8$ bpp.}
	\label{fig:depth_disp}
\end{figure*}

\begin{figure}[t]
\centering
\begin{minipage}{0.65\linewidth}
		\includegraphics[width=1\linewidth]{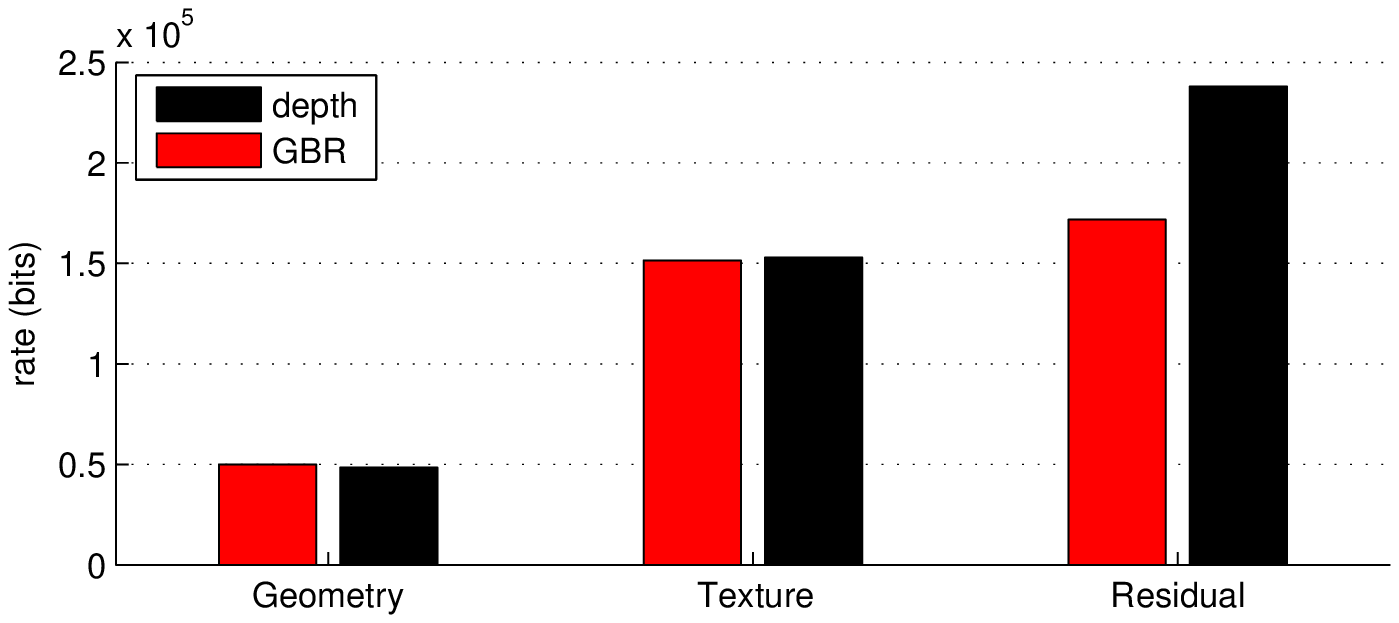}
		\centerline{(a)}
					\end{minipage}	
							\begin{minipage}{0.33\linewidth}
		\includegraphics[width=1\linewidth]{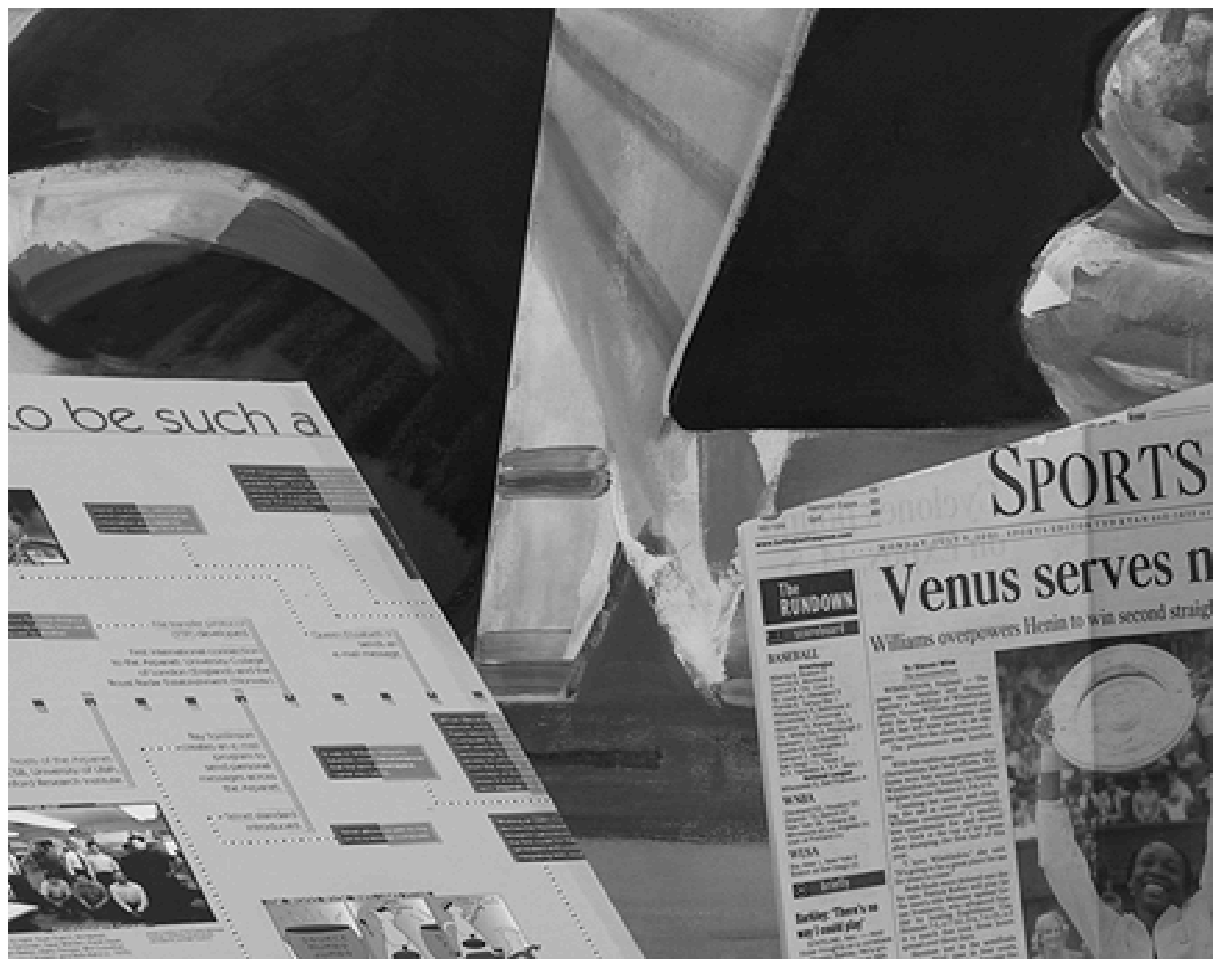}
		\centerline{(b)}
				\end{minipage}		
\caption{Coding rate distribution (a) between geometry, texture and residual components for GBR and depth-based representations. For both schemes, the average reconstruction quality of the $5$ views of \emph{venus}  dataset (b) is set to $32.93$ dB.}
	\label{fig:residualSize}
\end{figure}

\section{Multiview coding experiments}\label{sec:exp}

We now evaluate the performance of our novel GBR representation technique. We consider the multiview system described in Fig.~\ref{fig:framework} and show that i) the representation of the geometry with our graph-based approach leads to more efficient compression performance than depth-based schemes and ii) the graph-based representation of the geometry provides a better control of geometry coding artifacts than commonly used approaches for depth map compression. 

We first propose experiments where we measure the compressibility of the geometry signal in GBR and depth-based schemes in lossless representation scenarios, in the sense that view prediction is perfect. We focus on two views with only integer disparities in the ``squares 1" dataset introduced above. We build our GBR structure on the two first images. The reference image is not compressed and no residual is transmitted. Similarly, the depth-based scheme encodes one reference image, one depth image and the color residual of view $2$. The compression of the depth image is done with the lossless JPEG2000 codec \cite{jpeg2000}, while the luminance is also transmitted losslessly. We focus on the geometry rate only, and for both schemes the prediction is perfect. 
We first observe that the rate needed to compress the depth image is equal to $4.7$ kb, while the rate for graph information is equal to $2.2$ kb. Thus, the graph links provide a more compact description of the scene geometry than lossless compression of depth map images. Even though a more efficient technique could be considered for lossless depth compression, this first experiment shows that our graph obtains a good compressibility of its geometry signal, even in the lossless coding case. This case is however particular in the sense that lossless prediction only happens for very particular datasets. Moreover, coding schemes almost never operate in a lossless configurations.

We next evaluate performance in lossy compression scenarios. In natural images, losses are introduced because of a) non integer disparities or depth inaccuracies as shown in Sec.~\ref{sec:coding} and b) geometry compression with graph reduction or depth image compression in our GBR or depth-based scheme respectively. 
We study now the geometry compression artifacts. We use a more complex dataset  called ``squares 3" ($190\times190$) that contains more complex depth maps, since the foreground objects are not  parallel to the camera plane, unlike in the ``squares 1" and ``squares 2" datasets. The positions and the depths of the foregrounds are initialized randomly. Depth images corresponding to this new dataset are shown in Fig.~\ref{fig:exp3}. As the foreground objects do not have the same depth everywhere, new disoccluded pixels may appear. This is due to the fact that the foreground objects change size from one view to another. Since prediction algorithms simply project the pixels involved in a view, some additional pixels might be added to complete the view. They are handled by the residual images in the depth-based scheme, while they are simply added in the current graph level in our GBR. 
As in the previous experiment, we are interested in the geometry information compression only. We thus compress the geometry information with our GBR scheme, and compare this with the depth-based scheme where the depth image is encoded with JPEG2000. In both cases, we use the same encoding rate for the geometry information. As can be seen in Fig.~\ref{fig:exp3}~(c), the JPEG2000 depth compression leads to significant artifacts on the resulting depth maps and thus to high compensation error (Fig.~\ref{fig:exp3}~(e)).  With the GBR scheme, the geometry information is more accurate (Fig.~\ref{fig:exp3}~(d)), and the reconstruction results are better, as shown in Fig.~\ref{fig:exp3} (f). Similar observations can be made on natural sequence, as shown in Fig.~\ref{fig:depth_disp} for the ``sawtooth" dataset. We use the same comparison method and study the geometry compression artifacts. We compare the original depth map, the retrieved disparity map from the GBR and the compressed depth image (at similar bitrate). We observe that GBR provides better control over where to introduce losses and where to preserve geometry accuracy. More specifically, the reconstructed disparity map is piecewise constant but the edges are still sharp, in contrast to the approximation provided by JPEG2000 compression. Moreover, the level of geometry precision achieved by GBR is just enough to reconstruct the second viewpoint. We next show how these GBR properties lead to better reconstructed view quality.

We build another experiment with more images, \emph{i.e.}, a higher $N$,  and extend our study of the effect of geometry compression for the GBR and depth-based representations when, this time, reference, geometry and residual are coded. We run experiments where we represent the $5$ images of the ``venus" dataset (Fig.~\ref{fig:residualSize} (b)) using GBR and depth-based coding schemes. We select the coding total coding rates so that we achieve the same reconstruction quality ($32.93$ dB) while  geometry rates and color rates are kept similar in both coding schemes. In other words, we vary only the rate of the residual images. We show the rate distribution in Fig.~\ref{fig:residualSize} (a) for the two representations. We observe that for a constant geometry and texture rate, the depth-based scheme needs to send more residual information in order to achieve the same quality. In other words, the GBR has to perform less compensation after geometry compression, which means that it controls better the effect of geometry coding. While, similar observations are done at different target qualities, GBR gains with respect to depth-based scheme are highest at medium or high bitrates. 
The GBR, by the nature of its construction, cannot decrease its geometry rate below the minimum amount of information that is needed for one view prediction. Thus it looses its advantage with respect to depth-based representation in this rate range.

\begin{figure}[t]
\centering
		\includegraphics[width=0.8\linewidth]{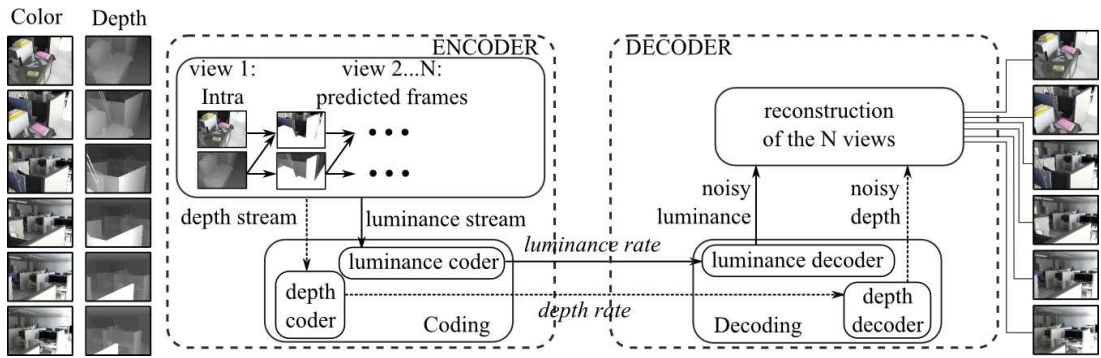}
	\caption{IPPPP depth-based multiview encoding scheme used for experiments.}
	\label{fig:disp}
\end{figure}

\begin{figure}[t]
\begin{minipage}{0.25\linewidth}
\center
		\includegraphics[width=1\linewidth]{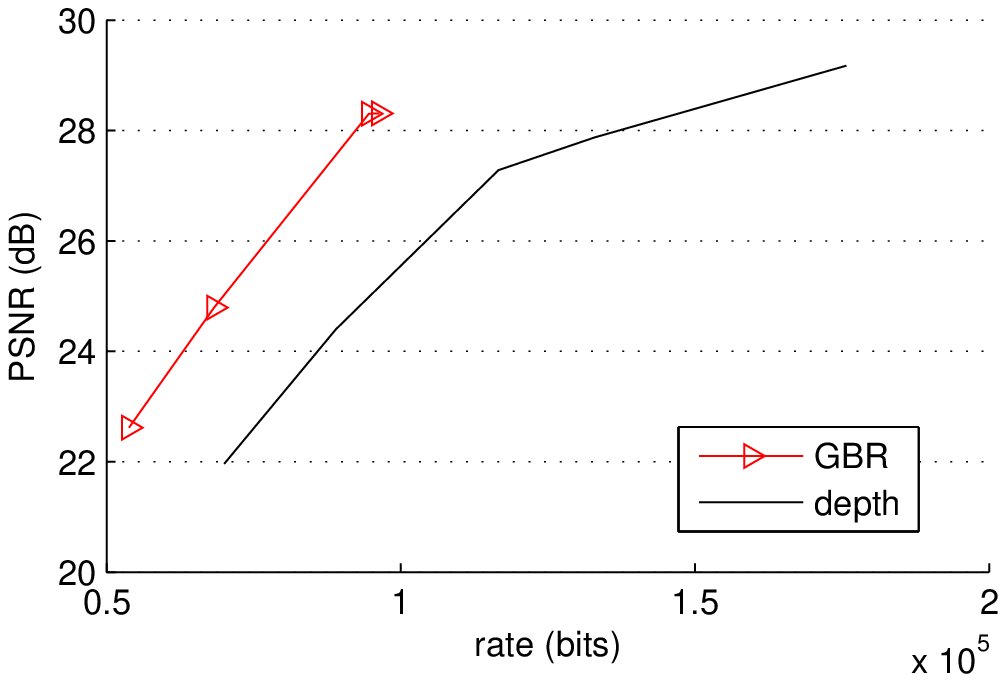}
		\centerline{(a)}
\end{minipage}
\begin{minipage}{0.39\linewidth}
\center
		\includegraphics[width=1\linewidth]{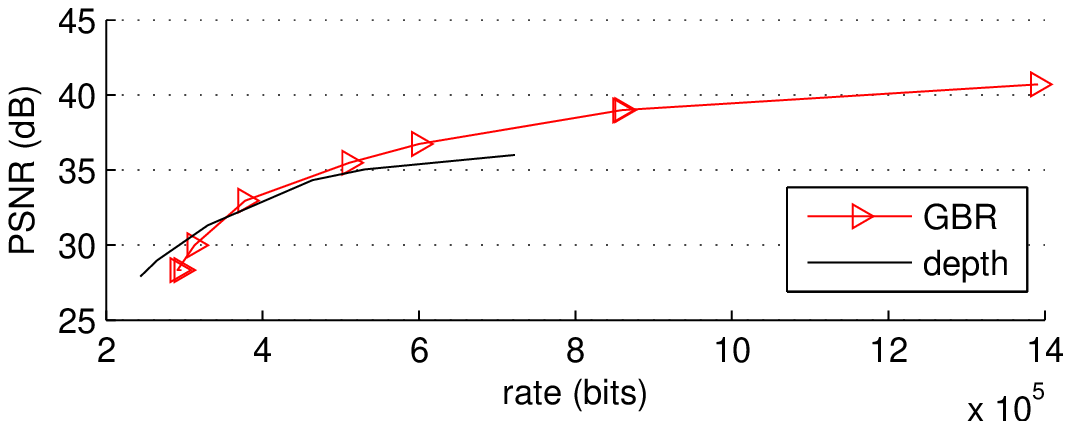}
		\centerline{(b)}
\end{minipage}
\begin{minipage}{0.36\linewidth}
\center
		\includegraphics[width=1\linewidth]{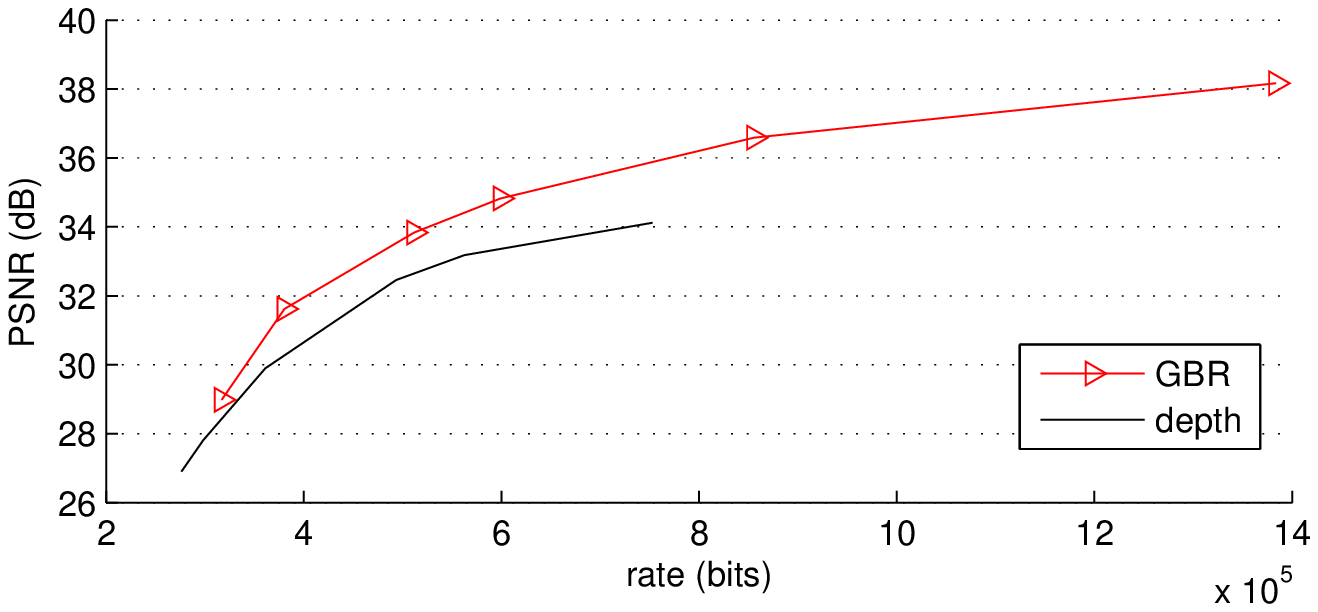}
		\centerline{(c)}
		\end{minipage}
	\caption{Rate-distortion performance comparisons between the GBR system and the depth-based scheme in a $IPPPP$ configuration for respectively (a) ``squares 2", (b) ``venus" and (c) ``sawtooth" test sequences.}
	\label{fig:exp4}	
\end{figure}

Finally, we present some rate-distortion (RD) performance evaluation results, where we compare the optimized GBR and optimized depth-based schemes in the scenario depicted in Fig.~\ref{fig:framework}. 
For the depth-based scheme, we consider a format of type IPPPPP, which means that a first view is transmitted along with its depth, and then, the other views are estimated iteratively by disparity-compensation using residual error data (for depth and color images). The block diagram of this scheme is illustrated in Fig.~\ref{fig:disp}. We build the GBR coder as explained before; it uses the same $N$ images (color and depth) as the depth-based scheme. The objective for both schemes is the reconstruction of $N$ color images. For both schemes, we simulate RD points at different quantization steps for geometry, color and residual compression. For the GBR scheme, we also vary the number of levels $L\leq N$ in the graph (the $N-L$ other levels are interpolated at the decoder side). In both schemes, we have distributed the rates of geometry, texture and residual optimally in order to maximize the reconstruction quality. In particular, we retain the convex envelope of these two RD point clouds in order to present the optimal RD curves for each scheme.
We present the  results obtained for the ``squares 2", ``venus" and ``sawtooth" datasets in Fig.~\ref{fig:exp4} (a), (b) and (c) respectively. We see that our scheme generally outperforms the depth-based approach. This is due to the fact that GBR controls the geometry compression, which leads to reduced residual error sizes.  We see however that, at low bitrates, the difference between the two schemes is smaller or that the depth approach is better for the ``Venus" dataset. The simple graph compression algorithm that we have designed is still limited when the bandwidth is too small. In particular, once we have removed all the intermediary images from the graph, we cannot reduce further the rate required for the geometry information in GBR. This fixed overhead leads to less competitive behavior at low rates. However, outside of the very low bitrate regime, the GBR representation leads to improved RD performance.

\section{Conclusion}
In this paper, we have proposed an alternative to depth-based representations for multiview image coding. Using graphs to describe connections between pixels of different views, our method manages to represent the geometry of the scene and to avoid the inter-view redundancies. At the same time, it increases the control on geometry compression artifacts in the reconstructed images. We have proposed a complete coding scheme based on this new graph-based representation and illustrated its potential in rate-distortion performance compared to depth-based schemes. Future work will focus on the development of more effective coding strategies in order to extend the performance of this promising GBR representation of multiview images. More precisely, we will investigate how GBR can handle non-integer disparity values, and quantization errors in other to improve the performance at low bitrate.

\ifCLASSOPTIONcaptionsoff
  \newpage
\fi

\section*{Acknowledgment}
This work has been partly supported by the  Hasler Foundation, within the project NORIA (novel image representation for future interactive multiview systems).

\bibliographystyle{IEEEtran}
\bibliography{/Users/thomasmaugey/Documents/EPFL/bibli/abbr,/Users/thomasmaugey/Documents/EPFL/bibli/epfl}

\begin{thebibliography}{10}
\providecommand{\url}[1]{#1}
\csname url@samestyle\endcsname
\providecommand{\newblock}{\relax}
\providecommand{\bibinfo}[2]{#2}
\providecommand{\BIBentrySTDinterwordspacing}{\spaceskip=0pt\relax}
\providecommand{\BIBentryALTinterwordstretchfactor}{4}
\providecommand{\BIBentryALTinterwordspacing}{\spaceskip=\fontdimen2\font plus
\BIBentryALTinterwordstretchfactor\fontdimen3\font minus
  \fontdimen4\font\relax}
\providecommand{\BIBforeignlanguage}[2]{{%
\expandafter\ifx\csname l@#1\endcsname\relax
\typeout{** WARNING: IEEEtran.bst: No hyphenation pattern has been}%
\typeout{** loaded for the language `#1'. Using the pattern for}%
\typeout{** the default language instead.}%
\else
\language=\csname l@#1\endcsname
\fi
#2}}
\providecommand{\BIBdecl}{\relax}
\BIBdecl

\bibitem{Alenya_G_2011_ieeesj}
G.~Alenya and C.~Torras, ``Lock-in time-of-flight ({TOF}) cameras: A survey,''
  \emph{IEEE Sensors Journal}, vol.~11, pp. 1917--1926, 2011.

\bibitem{Shum_HY_2003_tcsvt_sur_ibrct}
H.~Shum, S.~Kang, and S.~Chan, ``Survey of image-based representations and
  compression techniques,'' \emph{IEEE Trans. on Circ. and Syst. for Video
  Technology}, vol.~13, pp. 1020--1037, 2003.

\bibitem{Muller_K_2011_pieee_tdv_rudm}
K.~M\"uller, P.~Merkle, and T.~Wiegand, ``{3D} video representation using depth
  maps,'' \emph{Proc. IEEE}, vol.~99, no.~4, pp. 643--656, Apr. 2011.

\bibitem{Salvador_J_2013_mul_vrbfmcsr}
J.~Salvador and J.~Casas, ``Multi-view video representation based on fast monte
  carlo surface reconstruction,'' \emph{IEEE Trans. on Image Proc.}, vol.~22,
  pp. 3342--3352, 2013.

\bibitem{jmvm}
{ISO/IEC MPEG \& ITU-T VCEG}, ``Joint multiview video model ({JMVM}),''
  Marrakech, Morocco, Jan.13-19 2007.

\bibitem{Chai_JX_20000_psiggraph_ple_s}
J.~Chai, X.~Tong, S.~Chan, and H.~Shum, ``plenoptic sampling,'' in \emph{Proc.
  Int. Conf. on Computer graphics and interactive techniques}, 2000.

\bibitem{Kim_SY_2007_picip_mes_bdctdvuhdm}
S.~Kim and Y.~Ho, ``Mesh-based depth coding for {3D} video using hierarchical
  decomposition of depth maps,'' in \emph{Proc. IEEE Int. Conf. on Image
  Processing}, San Antonio, TX, USA, Sep. 2007.

\bibitem{Merkle_P_2007_picip_mvvpdrc}
P.~Merkle, A.~Smolic, K.~M\:uller, and T.~Wiegand, ``Multi-view video plus
  depth representation and coding,'' in \emph{Proc. IEEE Int. Conf. on Image
  Processing}, San Antonio, TX, US, Oct. 2007.

\bibitem{Yea_S_2009_jspic_vie_spmvc}
S.~Yea and A.~Vetro, ``View synthesis prediction for multiview video coding,''
  \emph{EURASIP J. on Sign. Proc.: Image Commun.}, vol.~24, pp. 89--100, 2009.

\bibitem{Tian_D_2009_pspie_vie_sttdv}
D.~Tian, P.~Lai, P.~Lopez, and C.~Gomila, ``View synthesis techniques for {3D}
  videos,'' \emph{Proc. of SPIE, the Int. Soc. for Optical Engineering}, vol.
  7443, 2009.

\bibitem{Muller_K_2013_tip_tdh_evcmvvdd}
K.~M\"uller, H.~Schwarz, D.~Marpe, C.~Bartnik, S.~Bosse, H.~Brust, T.~Hinz,
  H.~Lakshman, P.~Merkle, H.~Rhee, G.~Tech, M.~Winken, and T.~Wiegand, ``{3D}
  high-efficiency video coding for multi-view video and depth data,''
  \emph{IEEE Trans. on Image Proc.}, 2013.

\bibitem{Maugey_T_2013_picassp_gra_brcmg}
T.~Maugey, A.~Ortega, and P.~Frossard, ``Graph-based representation and coding
  of multiview geometry,'' in \emph{Proc. Int. Conf. on Acoust., Speech and
  Sig. Proc.}, Vancouver, Canada, 2013.

\bibitem{Maugey_T_2013_ivmsp_mul_icugba}
------, ``Multiview image coding using graph-based approach,'' in \emph{IEEE
  Workshop on 3D Image/Video Technologies and Applications (IVMSP)}, Seoul,
  Korea, Jun. 2013.

\bibitem{Morvan_Y_2007_ppcs_joi_dtbamvvc}
Y.~Morvan, D.~Farin, and P.~de~With, ``Joint depth/texture bit-allocation for
  multi-view video compression,'' in \emph{Picture Coding Symposium (PCS)},
  2007.

\bibitem{Kim_WS_2009_picip_dep_mdavrdc}
W.~Kim, A.~Ortega, P.~Lai, T.~D, and C.~Gomila, ``Depth map distortion analysis
  for view rendering and depth coding,'' in \emph{Proc. IEEE Int. Conf. on
  Image Processing}, Cairo, Egypt, Nov 2009.

\bibitem{Wang_Q_2012_tcsvt_fre_vvcrda}
Q.~Wang, X.~Ji, Q.~Dai, and N.~Zhang, ``Free viewpoint video coding with
  rate-distortion analysis,'' \emph{IEEE Trans. on Circ. and Syst. for Video
  Technology}, vol.~22, pp. 875–--889, 2012.

\bibitem{Cheung_G_2011_tip_dep_bamicdibr}
G.~Cheung, V.~Velisavlevic, and A.~Ortega, ``On dependent bit allocation for
  multiview image coding with depth-image-based rendering,'' \emph{IEEE Trans.
  on Image Proc.}, vol.~20, pp. 3179–--3194, 2011.

\bibitem{Rajei_B_2012_at_rat_damcdibrf}
B.~Rajei, T.~Maugey, and P.~Frossard, ``Rate-distortion analysis of multiview
  coding in a {DIBR} framework,'' \emph{Annals of Telecommunications}, 2013.

\bibitem{Liu_S_2011_tb_new_dctucv}
A.~Liu, P.~Lai, D.~Tian, and C.~Chen, ``New depth coding techniques with
  utilization of corresponding video,'' \emph{IEEE Trans. on Broadcasting},
  vol.~57, pp. 551--561, 2011.

\bibitem{Cheung_G_2011_mmsp_dep_mcugbttds}
G.~Cheung, W.~Kim, A.~Ortega, J.~Ishida, and A.~Kubota, ``Depth map coding
  using graph based transform and transform domain sparsiﬁcation,'' in
  \emph{IEEE Int. Workshop on Multimedia Sig. Proc.}, Hangzhou, China, Oct.
  2011.

\bibitem{Daribo_I_2012_picip_ari_ecassmpdvc}
I.~Daribo, G.~Cheung, and D.~Florencio, ``Arithmetic edge coding for
  arbitrarily shaped sub-block motion prediction in depth video coding,'' in
  \emph{Proc. IEEE Int. Conf. on Image Processing}, Orlando, FL, USA, Sep.
  2012.

\bibitem{Gelman_A_2012_tip_mul_icudloba}
A.~Gelman, P.~Dragotti, and V.~Velisavlevic, ``Multiview image coding using
  depth layers and an optimized bit allocation,'' \emph{IEEE Trans. on Image
  Proc.}, vol.~21, pp. 4092--4105, 2012.

\bibitem{Takyar_U_2013_spl_ext_ldirmn}
U.~Takyar, T.~Maugey, and P.~Frossard, ``Extended layered depth image
  representation in multiview navigation,'' \emph{accepted in Signal Processing
  Letters}, vol.~21, pp. 22--25, Jan. 2014.

\bibitem{Mao_Y_2013_picip_exp_hfdibrgbr}
Y.~Mao, G.~Cheung, A.~Ortega, and Y.~Ji, ``Expansion hole filling in
  depth-image-based rendering using graph-based interpolation,'' in \emph{Proc.
  IEEE Int. Conf. on Image Processing}, Vancouver, Canada, May 2013.

\bibitem{jpeg2000}
\BIBentryALTinterwordspacing
JPEG-2000, ``{ISO}/{IEC} {FCD} 15444-1: {JPEG} 2000 final comitee draft version
  1.0,'' 2000. [Online]. Available: \url{http://www.jpeg.org/FCD15444-1.htm}
\BIBentrySTDinterwordspacing

\end{thebibliography}


\end{document}